\documentclass[10pt,conference]{IEEEtran}
\usepackage{cite}
\usepackage{amsmath,amssymb,amsfonts}
\usepackage{algorithmic}
\usepackage{graphicx}
\usepackage{textcomp}
\usepackage{xcolor}
\usepackage[hyphens]{url}
\usepackage{fancyhdr}
\usepackage{hyperref}
\usepackage{xspace}
\usepackage{enumitem}
\usepackage{tikz}
\usepackage{colortbl}
\usepackage{subfigure}
\usepackage{pifont}
\usepackage[numbers,sort&compress]{natbib}
\usepackage[linesnumbered,ruled,vlined]{algorithm2e}

\def\hpcacameraready{}

\newcommand\hpcaauthors{Zhen Liu$\dagger$*
, Wenzhe Zhu$\dagger$*, Yongkun Li$\dagger$,Yinlong Xu$\dagger$}
\newcommand\hpcaaffiliation{University of Science and Technology of China$\dagger$}
\newcommand\hpcaemail{}
\newcommand{\hpcayear}{}

\def\aeopen{} 

\newcommand*\bcircled[1]{\tikz[baseline=(char.base)]{
            \node[shape=circle,fill,inner sep=0pt] (char) {\textcolor{white}{#1}};}}

\renewcommand{\paragraph}[1]{\noindent {\bf #1}}
\newcommand\sysname{\textsf{FOCUS}\xspace}
\newcommand*\circled[1]{\tikz[baseline=(char.base)]{
            \node[shape=circle, fill=black, text=white, inner sep=.5pt] (char) {#1};}}

\newcommand\PBRB{SeaCache\xspace}
\newcommand\PLOG{PLog\xspace}
\newcommand\TLOG{PLog\xspace}
\newcommand\DLOG{DLog\xspace}
\newcommand\CLOG{CLog\xspace}
\newcommand{\code}[1]{\texttt{#1}}
\newcommand{\bfcode}[1]{\textsf{#1}}
\newcommand{\italic}[1]{\textit{#1}}

\title{\sysname: Boosting Schema-aware Access for KV Stores via Hierarchical Data Management}
\author{
{\rm Yongkun Li$^1$, Zhen Liu$^1$, Patrick P. C. Lee$^2$, Jiayu Wu$^1$, Yinlong Xu$^{1,3}$}\\
{\rm Yi Wu$^4$, Liu Tang$^4$, Qi Liu$^4$, Qiu Cui$^4$}
\vspace{3pt}\\
\textit{$^1$University of Science and Technology of China}\quad
\textit{$^2$%College of Computer Science,
The Chinese University of Hong Kong} \\ \textit{$^3$Anhui Province Key
Laboratory of High Performance Computing, USTC \quad $^4$PingCAP} }

\author{
  \ifdefined\hpcacameraready
    \IEEEauthorblockN{\hpcaauthors{}}
      \IEEEauthorblockA{
        \hpcaaffiliation{} \\
        \hpcaemail{}
      }
  \else
    \IEEEauthorblockN{\normalsize{HPCA \hpcayear{} Submission
      \textbf{\#\hpcasubmissionnumber{}}} \\
      \IEEEauthorblockA{
        Confidential Draft \\
        Do NOT Distribute!!
      }
    }
  \fi 
}

\fancypagestyle{camerareadyfirstpage}{%
  \fancyhead{}
  
  \fancyhead[C]{
    \ifdefined\aeopen
    {\hpcayear{} }    
    \else
      \ifdefined\aereviewed
      \parbox[][12mm][t]{13.5cm}{\hpcayear{} IEEE International Symposium on High-Performance Computer Architecture (HPCA)}
      \else
      \ifdefined\aereproduced
      \parbox[][12mm][t]{13.5cm}{\hpcayear{} IEEE International Symposium on High-Performance Computer Architecture (HPCA)}
      \else
      \parbox[][0mm][t]{13.5cm}{\hpcayear{} IEEE International Symposium on High-Performance Computer Architecture (HPCA)}
    \fi 
    \fi 
    \fi 
    \ifdefined\aeopen 
    \fi 
    \ifdefined\aereviewed
      \includegraphics[width=12mm,height=12mm]{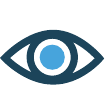}
    \fi 
    \ifdefined\aereproduced
      \includegraphics[width=12mm,height=12mm]{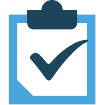}
    \fi
  }
  \fancyfoot[C]{}
}
\begin{document}
\maketitle

%Enables the camera ready header and footer
\ifdefined\hpcacameraready 
  \thispagestyle{camerareadyfirstpage}
  \pagestyle{empty}
\else
  \thispagestyle{plain}
  \pagestyle{plain}
\fi

\newcommand{\hpcaheight}{0mm}
\ifdefined\eaopen
\renewcommand{\hpcaheight}{12mm}
\fi

%%%%%%%%%%%%%%%%%%%%%%%%%%%%%%%%%%%%%%%%
%%%%%%%% -- PAPER CONTENT STARTS -- %%%%%%%%%

\begin{abstract}
% Persistent key-value (KV) stores are crucial for data-intensive applications such as graph stores, OLTP/OLAP databases, and file systems. Recent advancements in Non-Volatile Memory (NVM) have focused on enhancing KV stores due to NVM's lower latency and higher bandwidth. However, the integration of KV stores within broader application architectures remains problematic due to the mismatch between structured data models in applications and the flat key-value model of KV stores. Current solutions typically introduce an intermediary software layer for semantic translation, leading to I/O Amplification and I/O Splitting, which undermine the performance benefits of NVM.

% To this end, we propose \textsf{FOCUS}, a log-structured KV store optimized for fine-grained hierarchical data management and schema-aware access.
% \textsf{FOCUS} employs a hierarchical data management, decomposing KV pairs into finer fields based on user specifications to enable \emph{schema-aware access}, providing native support for selective reads/updates to fields within KV pairs.
% Such \emph{schema-aware access} feature eliminates the I/O Amplification and I/O Splitting cause by the intermediary layer, thereby fully leveraging NVM's advantages.

% Evaluation results show that \textsf{FOCUS} outperforms existing NVM-optimized KV stores on schema-based workloads (by 2.0$\times$ to 2.8$\times$). \textsf{FOCUS} is also scalable to the number of threads and robust to tunable parameters.

Persistent key-value (KV) stores are critical infrastructure for data-intensive applications.
Leveraging high-performance Non-Volatile Memory (NVM) to enhance KV stores has gained traction. However, previous work has primarily focused on optimizing KV stores themselves, without adequately addressing their integration into applications. Consequently, existing applications, represented by NewSQL databases, still resort to a flat mapping approach, which simply map structured records into flat KV pairs to use KV stores. Such semantic mismatch may causes significant I/O amplification and I/O splitting under production workloads, harming the performance. To this end we propose \sysname, a log-structured KV store optimized for fine-grained hierarchical data organization and schema-aware access. \sysname introduces a hierarchical KV model to provide native support for upper-layer structured data. We implemented \sysname from scratch. Experiments show that \sysname can increase throughput by 2.1$\sim$5.9$\times$ compared to mainstream NVM-backed KV stores under YCSB SQL workloads.

\end{abstract}
\section{Introduction}\label{sec:intro}
% Key-value stores (KVSs) have found wide application in modern software systems.
Persistent key-value (KV) stores have become fundamental infrastructure for supporting data-intensive applications, including graph stores~\cite{Da15FlashGraph,Aapo23GraphChi}, OLTP/OLAP databases~\cite{Myrocks16,Huang2020tidb,YugabyteDB17,Cao22polardb,Corbett13spanner,pilman2017fastscan,caow2020polardb,jiang2020hologres}, and file systems~\cite{Aghayev19file,Harter4analysis}. In recent years, utilizing Non-Volatile Memory (NVM) to enhance KV stores has gained significant traction. NVM not only features data persistence, byte-addressability, and high density~\cite{Duan2021hardware, Dulloor2016data}, but also offers up to 100$\times$ lower access latency and 10$\times$ higher bandwidth than SSDs~\cite{Yao2020matrixkv, Kannan2018redesigning}. These merits greatly facilitate the construction of high-speed, persistent KV stores. However, the properties of NVM dictate a pronounced demand for fine-grained access, sequential read/writes, and low software overhead to unleash its potential~\cite{Yang20empirical, izraelevitz2019basic, Gotze2020data}.

To accommodate these characteristics in KV stores, extensive research has been conducted. One category of study, which we term NVM-based enhancement~\cite{Kaiyrakhmet2019slmdb, Kim2022listdb, Yao2020matrixkv, Zhang2021chameleondb}, maintains the established Log-structured Merge-Tree (LSM-Tree) architecture while judiciously adapting it to NVM traits, thus mitigating long-standing issues of read/write amplification and write stall. Another category, which we refer to as NVM-oriented redesign~\cite{Benson2021viper, Cai2023bonsaikv, Chen2020flatstore}, involves a more comprehensive overhaul in architecture, data organization, and workflow of KV stores, specifically engineered for NVM to capitalize on its features. Despite distinct methodologies, both categories have demonstrated promising performance gains.

Despite extensive NVM optimizations that focused solely on KV stores themselves, the efficient integration of KV stores into applications presents a subtle challenge. Structured records from applications must be mapped into KV pairs for storage. Take, as an example, the state-of-the-art NewSQL databases~\cite{Taft20cockroachdb,Corbett13spanner,Huang2020tidb,YugabyteDB17}, which operate on table rows (a.k.a., records) for intuitive handling. They primarily adopt either a consolidated mapping, which encapsulates all attributes of a record into a single KV pair, or a scattered mapping, which distributes these attributes as separate KV pairs. We collectively refer to them as \emph{flat data mapping}, as structured records are uniformly mapped into flat, indivisible KV pairs.

\begin{figure}[!t]
\centering
\subfigure[I/O Amplification]{
        \includegraphics[width=0.4\linewidth]{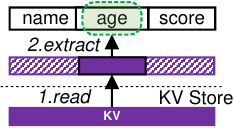}
        \label{fig:consolidated_inefficiency}
}
% \hspace{-10pt}
\subfigure[I/O Splitting]{
       \includegraphics[width=0.40\linewidth]{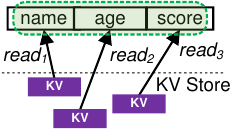}
        \label{fig:scattered_inefficiency}
}
   % \vspace{-10pt}
	\caption{\textbf{Inefficiency of flat data mapping.} we omit the primary key here.}
	\label{fig:flat_mapping_inefficiency}
	\vspace{-10pt}
\end{figure}

Both our analysis and experiments demonstrate the inefficiency of flat data mapping in NVM environments. Specifically, the consolidated mapping method results in I/O amplification (see Figure~\ref{fig:consolidated_inefficiency}), as accessing a single attribute requires handling the entire record due to the indivisible nature of the KV pair. Conversely, the scattered mapping method leads to I/O splitting (see Figure~\ref{fig:scattered_inefficiency}), where attributes within a record are managed as separate KV pairs, necessitating numerous small-sized KV operations for cross-attribute accesses, such as full-record reads. Consequently, flat data mapping fails to satisfy the requirements for sequential access, low software overhead, and fine-grained access, which are crucial for exploiting NVM's characteristics. During our experiments, we observed a 1.7$\sim$3.6$\times$ latency increase for I/O amplification and an 3.3$\sim$4.8$\times$ increase for I/O splitting, compared to the ideal (see \S\ref{subsec:limitations}).
% Unfortunately, our research indicates that such a flat mapping approach results in significant performance bottlenecks within NVM environments, impeding the transmission of KV stores' high performance to applications.
% While upper-layer applications often employ structured data records for intuitive data handling (e.g., table rows in NewSQL databases~\cite{Huang2020tidb, Corbett13spanner, Taft20cockroachdb} and vertices or edges in graph stores~\cite{}), KV stores only provide a primitive interface of KV pairs. To bridge this disparity, an intermediary \emph{flat data mapping} is traditionally employed to convert structured records into flat and indivisible KV pairs for storage. 

Our insight is that, instead of adhering to a flat KV abstraction, a promising strategy is to incorporate a novel \emph{hierarchical data management}. On one hand, each KV pair is decomposed into finer-grained fields, allowing for independent updates and reads on these fields to enhance flexibility. On the other hand, fields belonging to the same KV pair are cohesively managed to improve data locality, simplify indexing, and minimize metadata overhead. By mapping a record to such a hierarchical KV pair and aligning the schema (i.e., the attributes in the record) with the fields within the KV pair, the upper-layer NewSQL database can selectively read or update desired attributes by operating on the corresponding fields, thus eliminating the I/O amplification or I/O splitting imposed by flat data mapping. We term this capability \emph{schema-aware access}. Notably, without explicit schema specification, KV pairs retain a flat structure, ensuring backward compatibility.

In this paper, we realize fine-grained data management and schema-aware access in \sysname, a log-structured KV store optimized for \underline{f}ine-grained hierarchical data \underline{o}rganization \underline{c}oupled with \underline{s}chema-aware access. \sysname preserves the simplicity and compatibility of the original KV interface while offering essential semantic support for NewSQL databases to incorporate schema information. By adopting a schema-friendly layout, \sysname organizes schema metadata and persistent data across DRAM and NVM in a coordinated manner, facilitating efficient field parsing. For data persistence, \sysname stores KV pairs and their corresponding fields in a persistent log (\PLOG) to maintain the advantage of sequential writes. \PLOG is structured into two sub-logs: \CLOG and \DLOG. \CLOG captures full updates on each KV pair, whereas \DLOG tracks delta updates involving specific fields within KV pairs. This separation of updates, which have distinct lifetimes, into \CLOG and \DLOG, significantly reduces data migration during garbage collection. Moreover, to effectively manage fields within the same KV pair, \sysname employs a sequential-write ahead of in-place merge (swim) mechanism, which asynchronously applies delta updates from \DLOG to \CLOG, thereby constructing the latest complete state of KV pairs. Lastly, to mitigate the inevitable small I/Os resulting from reads on specific tiny fields, \sysname employs a schema-aware cache (SeaCache) that retains hot KV pairs in memory to accelerate read operations. This cache is integrated with the global index management, thereby preventing the latency associated with ineffective probes during cache misses.
Our contributions are summarized as follows.
\begin{itemize}[leftmargin=*]
\item
We analyze three state-of-the-art data mapping strategies adopted by mainstream NewSQL databases, including TiDB~\cite{Huang2020tidb}, Bigtable~\cite{Bigtable15}, and YugabyteDB~\cite{YugabyteDB17}, and reveal their performance limitations due to mapping records to flat, indivisible KV pairs.
\item
We design \sysname, which realizes hierarchical data management and supports schema-aware access, through (\romannumeral 1) a simple yet flexible API for operating hierarchical KV pairs, allowing users to specify schema information and perform schema-aware access, (\romannumeral 2) an efficient two-layer persistent log that stores full updates and delta updates of KV pairs separately, where delta updates are asynchronously applied to construct the consistent version of KV pairs while maintaining a compact data layout. (\romannumeral 3) a schema-aware cache to accelerate reads on specific tiny fields.
\item
We implemented \sysname from scratch and evaluated it against several state-of-the-art NVM-powered KV stores. Experiments show that \sysname outperforms Pmem-RocksDB~\cite{Pmem-RocksDB} and ListDB~\cite{Kim2022listdb} with two common mapping strategies. For instance, \sysname achieves 3.5-5.9 $\times$ better performance in partial reads and updates with consolidated mapping (see \S\ref{subsec:flat}). Compared to scattered mapping (see \S\ref{subsec:flat}), \sysname improves performance by up to 3.1$\times$. Additionally, due to schema metadata separation, \sysname maintains an up to 2.1$\times$ advantage even under full access, owing to reduced value parsing overhead.
\end{itemize}

% We will release our source code in the final paper.
\section{Background}
\label{sec:background}

\subsection{Persistent KV Stores and Applications}
\label{subsec:kvs}

\noindent\textbf{Persistent KV Stores.}
Persistent key-value (KV) stores, abbreviated as KV stores, abstract data into collections of KV pairs and provide users with an intuitive interface for data manipulation. Basic operations include \texttt{GET}, \texttt{PUT}, \texttt{SCAN}, and \texttt{DELETE}. Specifically, \texttt{GET} retrieves the value associated with a given key, \texttt{PUT} inserts or updates a KV pair, \texttt{SCAN} traverses a range of KV pairs, and \texttt{DELETE} removes a KV pair by the given key. This simplified data model eliminates complex schemas and relationships among data, enhancing storage efficiency and scalability. 
Modern KV stores often combine an append-only Log-Structured Merge-Tree (LSM-Tree)~\cite{O1996log} to boost performance. LSM-Tree improves write performance by buffering KV pairs in an in-memory \emph{MemTable} before flushing them to HDDs/SSDs as \emph{SSTables} in the $L_0$ level, thus transforming random writes into sequential ones. In addition, it periodically compacts data into higher levels ($L_1$, $L_2$, ..., $L_n$) to balance read performance and space efficiency.

\noindent\textbf{NewSQL databases.}
To leverage the advantages of KV stores, data-intensive applications are increasingly adopting them as underlying storage~\cite{Da15FlashGraph,Aapo23GraphChi, Myrocks16,Huang2020tidb,YugabyteDB17,Cao22polardb,Corbett13spanner}. NewSQL databases exemplify this trend \cite{Huang2020tidb, Corbett13spanner, Taft20cockroachdb}.  
These databases retain the SQL querying capabilities of traditional relational databases (RDBMS) while incorporating the high performance and scalability of KV stores.
This synthesis makes NewSQL databases a compelling choice for modern applications that require both stringent data consistency and the ability to efficiently manage large-scale, distributed workloads. Examples include financial transactions~\cite{Oracle95,Huang2020tidb,yang2022oceanbase}, e-commerce platforms~\cite{Corbett13spanner,Cao22polardb,DynamoDB12}, and real-time analytics~\cite{pilman2017fastscan,caow2020polardb,jiang2020hologres}.

\subsection{Non-volatile Memory (NVM).}
\label{subsec:nvm_perf_characteristic}
% Rising demands for fast data storage and processing are increasingly challenging traditional storage devices. Fortunately, innovations in storage media technologies have introduced Non-volatile Memory (NVM) as a promising solution.

\noindent\textbf{Characteristics.}
NVM combines several unique characteristics that make it particularly suitable for modern data-intensive applications and storage systems:
\begin{itemize}[leftmargin = *]
    \item \emph{Persistence}: Same as block devices, e.g., HDDs or SSDs, NVM retains data after power loss, ensuring durability..
    \item \emph{Low latency}: NVM provides significantly lower latency than HDDs and SSDs, approaching the speed of DRAM.
    \item \emph{Byte-addressability}: Unlike block devices, NVM allows for byte-addressable access, enabling precise data operations.
\end{itemize}

\noindent\textbf{Storage media.}
NVM encompasses various foundational storage media, including but not limited to 3D XPoint~\cite{Hady2017platform}, resistive RAM (ReRAM)~\cite{Akinaga2010resistive}, STT-magnetoresistive RAM (MRAM)~\cite{Tulapurkar2005spin}, phase-change memory (PCM)~\cite{Raoux2008phase}, and ferroelectric RAM (FRAM)~\cite{kato2007overview}. 
Among these, Intel's 3D XPoint is a prominent example. 
% It features a unique cross-point architecture with memory cells located at the intersections of perpendicular word lines and bit lines, allowing for high density and fast data access without the use of transistors. 
Based on this technology, Intel developed well-known Optane DC Persistent Memory (DCPMM) and released it in 2019. 
Although DCPMM was announced to be discontinued in 2022, the industry has continued to develop NVM products based on alternative storage media. Successful commercial technologies include NRAM~\cite{Gilmer2018nram}, battery-backed DRAM~\cite{Mehra2004fast}, and PRAM~\cite{Kim2005reliability}. Also, NVM remains a hot research topic, with numerous relevant papers still emerging.

\subsection{NVM-powered KV Stores.}
\label{subsec:existing}
\noindent\textbf{Key considerations.}
Due to NVM's unique characteristics, merely replacing block-based storage with an NVM-based one does not achieve the best performance. Extensive research has explored optimal access patterns and system architectures for NVM~\cite{Yang20empirical, izraelevitz2019basic, Gotze2020data}. In essence, the design of storage systems, including KV stores, should adhere to the following principles:

\noindent\textbf{\bcircled{1}}\emph{ Sequential access.} 
The low latency of NVM are best utilized through sequential access patterns~\cite{Yang20empirical, izraelevitz2019basic, Gotze2020data}. Thus, torage systems should be engineered to facilitate sequential access.

\noindent\textbf{\bcircled{2}}\emph{ Low software overhead.} 
The microsecond-level latency of NVM can be significantly undermined by excessive software overhead, e.g., serialization and deserialization costs~\cite{Kannan2018redesigning}. Therefore, it's crucial to streamline the software overhead.

\noindent\textbf{\bcircled{3}}\emph{ Fine-grained access.}
To leverage the byte-addressability of NVM, storage systems need to precisely handle fine-grained operations, thus minimizing unnecessary data access volume.

\noindent\textbf{Existing designs.}
Based on their methodologies to satisfy the above requirements, existing NVM-powered KV store designs can be divided into two categories, which we term \emph{NVM-based enhancement} and \emph{NVM-oriented redesign} respectively.

NVM-based enhancement~\cite{Kaiyrakhmet2019slmdb, Kim2022listdb, Yao2020matrixkv, Zhang2021chameleondb} retains LSM-Tree architecture since its append-only feature naturally yields sequential access. NVM is then employed to enhance fine-grained access and mitigate the inherent high software overhead of LSM-Tree.
% For instance, MatrixKV~\cite{Yao2020matrixkv} leverages NVM to perform more cost-effective and fine-grained compaction for $L_0$ and $L_1$ levels to address write stall (WS), and reduce LSM-tree depth to mitigate write amplification (WA). 
For example, ListDB~\cite{Kim2022listdb} employs Index-Unified Logging (IUL) to write each KV pair to NVM only once for persistence. By leveraging NVM’s byte-addressability, it lazily converts IUL entries into SkipList elements, masking the overhead of logging and \emph{MemTable} flushes, thereby resolving Write Stall.
% SLM-DB~\cite{} uses NVM to store the MemTable and Immutable MemTable, thereby eliminating the need for write-ahead logging (WAL). 
% By maintaining a single level of SSTables, it avoids compaction, significantly reducing write amplification (WA).

In contrast, NVM-oriented redesign~\cite{Benson2021viper, Cai2023bonsaikv, Chen2020flatstore} involves a more comprehensive overhaul to provide more native adaptation for NVM. A representative approach is to replace the LSM-Tree with a single-layer persistent log. This lightweight architecture retains the append-only feature while circuvmenting the high overhead of the LSM-Tree.
For example, FlatStore~\cite{Chen2020flatstore} batches index metadata and small-sized KV pairs in a persistent log to create a sequential access pattern favored by NVM, and combines a partitioned hash table and a global Masstree~\cite{Mao2012cache} for efficient indexing. 
% Similarly, BonsaiKV~\cite{Cai2023bonsaikv} uses a log for persistence and an in-DRAM Masstree for indexing, while proposing techniques to further address congestion in NVM.
\section{Hierarchical Data Management}
\label{sec:management}

To motivate hierarchical data management, we first examine existing KV store integration in NewSQL databases, which relies on flat data mapping (\S\ref{subsec:flat}), and its limitations in mapping structured records into flat, indivisible KV pairs in the context of NVM (\S\ref{subsec:limitations}). We then present the idea of hierarchical data management and associated challenges in system design (\S\ref{subsec:idea}).

\subsection{KV Store Integration Through Flat Data Mapping.}
\label{subsec:flat}
\begin{figure}[!t]
\centering
\subfigure[Consolidated Mapping]{
        \includegraphics[width=0.47\linewidth]{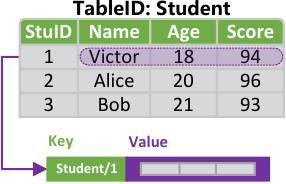}
        \label{fig:consolidated_mapping}
}
% \hspace{-10pt}
\subfigure[Scattered Mapping]{
       \includegraphics[width=0.44\linewidth]{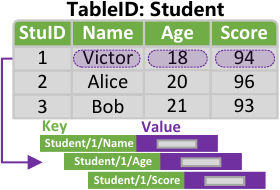}
        \label{fig:scattered_mapping}
}
   % \vspace{-10pt}
	\caption{\textbf{Flat data mapping strategies.} the primary key is StuID, which is encoded into the key, thus do not need to be stored in the value.}
	\label{fig:flat_mapping_strategies}
	\vspace{-10pt}
\end{figure}

% However, despite extensive NVM optimizations focused exclusively on KV stores themselves (see \S\ref{subsec:existing}), their integration within broader application architectures relies on a flat mapping to map structured data into plain KV pairs.
\noindent\textbf{Flat data mapping.}
NewSQL databases retains the structured tabular format of RDBMS, allowing users to perform create, read, update, and delete (CRUD) operations on table rows (a.k.a. records).
Each record consist of a primary key and multiple other attributes.
To leverage KV stores in NewSQL databases, flat data mapping is employed to transform structured records into flat KV pairs for management. Representative mapping strategies include:

\begin{itemize}[leftmargin=*]
\item
\emph{Consolidated mapping}: It encapsulates all attributes of a record into one KV pair, with the key encoded as "table\_id/primary\_key" (see Figure~\ref{fig:consolidated_mapping}). Here, table\_id identifies the corresponding table, and primary\_key identifies the record. MyRocks~\cite{Myrocks16} and TiDB~\cite{Huang2020tidb} utilize this approach.
\item
\emph{Scattered mapping}: It stores each attribute of a record as a separate KV pair, with the key encoded as "table\_id/primary\_key/attribute\_name" (see Figure~\ref{fig:scattered_mapping}). Here, attribute\_name refers to the name of the attribute being stored. Bigtable~\cite{Bigtable15}, HBase~\cite{HBase}, and YugabyteDB~\cite{YugabyteDB17} adopt this approach.
\item
\emph{Hybrid mapping}: It requires users to specify which attributes should be separated, as in \emph{scattered mapping}, and which should be grouped together, similar to \emph{consolidated mapping}. The grouped attributes are indexed by the key "table\_id/primary\_key/family\_name", where family\_name refers to the identifier of the attribute group. ScyllaDB~\cite{ScyllaDB15} and CockroachDB~\cite{Taft20cockroachdb} adopt this approach.
\end{itemize}

% \begin{figure}[!t] 
% 	\centering
% 	\includegraphics[width=1.0\linewidth]{figures/paper_figs/kv_mapping.pdf}
 
%     \vspace{-10pt}
% 	\caption{Two mapping approaches: consolidated and scattered conversion.}
% 	\label{fig:background_kv_mapping}
%      \vspace{-10pt}
% \end{figure}

\noindent\textbf{Modeling and analysis.}
Mathematically, flat data mapping can be modeled as a set partitioning problem. Given a set of \(N\) attributes, it can be divided into arbitrary non-empty, mutually exclusive, and collectively exhaustive subsets, where attributes within a subset would be packed into the same KV pair, while attributes from different subsets are placed into separate KV pairs.
Our goal is to align the partition with the \emph{attribute co-access pattern} by (\romannumeral 1) grouping co-accessed attributes into the same subset, thereby consolidating them into the same KV pair, and (\romannumeral 2) distributing separately accessed attributes into different subsets, thus placing them in distinct KV pairs.
Meeting this goal ensures that the attributes involved in an access match exactly to a specific subset within the partition, which means that (\romannumeral 1) all necessary attributes are included in the subset, and thus in the corresponding KV pair, allowing the operation to be completed in a single step, and (\romannumeral 2) no extra attributes are included, minimizing unnecessary data access. This perfectly aligns with the characteristics of NVM.
% For example, consider a student table with attributes \{\emph{name}, \emph{age}, \emph{major}, \emph{GPA}, \emph{hobby}\}. If the attribute co-access pattern is represented by the partition \{\{\emph{name}, \emph{age}\}, \{\emph{major}, \emph{GPA}\}, \{\emph{hobby}\}\}, indicating that attributes within each subset are frequently accessed together, then \emph{name} and \emph{age} should be packed into the first KV pair, \emph{major} and \emph{GPA} into the second KV pair, and \emph{hobby} should be stored as a separate KV pair.

With this model, we can evaluate the aforementioned mapping strategies and identify their most suitable attribute co-access patterns. For example, consolidated mapping partitions all attributes into a single subset, performing best when all attributes are accessed together. Conversely, scattered mapping divides \(N\) attributes into \(N\) subsets, each with one attribute, excelling when attributes are accessed individually. Finally, hybrid mapping spans the spectrum between these extremes, but requiring the user to fully understand the attribute co-access pattern and decide the partitioning scheme accordingly.

Theoretically, for an attribute set, the above mapping strategies cover all possible partitions. Therefore, provided the attribute co-access pattern is identified and fixed, i.e., for any pair of attributes, whether they would be co-accessed or separately accessed is deterministic or nearly deterministic, we can determine a matching partitioning scheme for the attribute set, and thus, the ideal data mapping approach.
% defining what is a attribute co-access pattern.

\subsection{Limitations}
\label{subsec:limitations}
As noted in previous studies~\cite{Cooper10ycsb,TPCC}, real-world workloads typically comprise a mix of the following operations. Unfortunately, we find their combination generates a volatile and dynamic pattern in co-accessing attributes, leading to unavoidable misalignment with any static partitioning scheme of the attribute set, thus rendering flat data mapping ineffective.
% Unfortunately, in real-world SQL workloads, requests with distinct attribute co-access patterns are frequently mixed together. This results in inevitable misalignment with any fixed partitioning of the attributes, rendering flat data mapping ineffective.
% s noted in previous studies~\cite{Cooper10ycsb,TPCC}
\begin{itemize}[leftmargin=*]
\item
\texttt{INSERT}: Insert a new record, often with multiple attributes.
\item
\texttt{UPDATE}: Updating the value of a single attribute in a record
\item
\texttt{READ}:  Read a record, either a randomly chosen single attribute or all attributes.
\item
\texttt{SCAN}: Sequentially scan a randomly chosen number of records, beginning with a specified record ID.
\end{itemize}
In terms of the range of accessed attributes, these operations fall into two distinct patterns: \emph{full access} and \emph{partial access}.
An \texttt{INSERT} operation involves full access, as all attributes of a record are inserted. An \texttt{UPDATE} operation represents partial access, modifying only one attribute. \texttt{READ} and \texttt{SCAN} operations can be either full or partial access, depending on user specifications.
Evidently, full access and partial access present irreconcilable conflicts in ideal partitioning schemes. Full access benefits from grouping all attributes into the same subset, while partial access prefers each attribute to be assigned to a separate subset. Simultaneously accommodating both access patterns is paradoxical.

To illustrate, consider using consolidated mapping. While full access operations can be conveniently served, partial access operations suffer significantly \emph{I/O amplification}. This is because even updating or reading a single attribute still requires handling the entire record due to the indivisible nature of KV pairs. Moreover, the software overhead for serializing and deserializing the entire record is unamortizable during partial access. This overhead, negligible in traditional slow block device environments, becomes considerable with the microsecond-level latency of NVM.
Considering the trend of wide-column tables, where a record often has nearly a hundred attributes~\cite{Bigtable15}, the I/O amplification and software overhead from partial access can reach approximately 100$\times$ in real-world workloads. This contradicts the \textbf{\bcircled{3}} fine-grained access and \textbf{\bcircled{2}} low software overhead principles demanded by NVM.

Regarding scattered mapping, although I/O amplification is avoided by storing attributes as separate KV pairs, allowing precise partial access, this approach suffers \emph{I/O splitting} during full access. Since each KV pair corresponds to a single attribute, a full access operation, such as reading an entire record, must be divided into multiple KV commands. Given that an attribute typically has a small size, often only tens of bytes, this results in numerous small-sized I/Os detrimental to NVM performance and violating the \textbf{\bcircled{1}} sequential access principle.
Moreover, scattering records into tiny KV pairs leads to poor data locality and an explosion in the number of KV pairs, which increases management overhead, thus violating the \textbf{\bcircled{2}} low software overhead principle.

\begin{figure}[!t]
        \centering
        \includegraphics[width=0.8\linewidth]{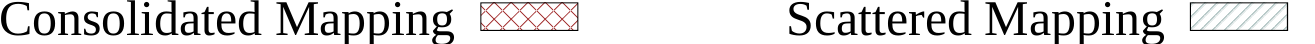}
\subfigure[Full access]{
        \includegraphics[width=0.4\linewidth]{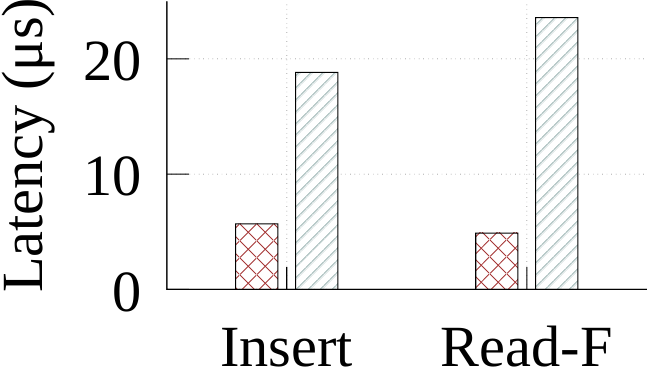}
        \label{fig:full_access_breakdown}
}
% \hspace{-10pt}
\subfigure[Partial access]{
       \includegraphics[width=0.4\linewidth]{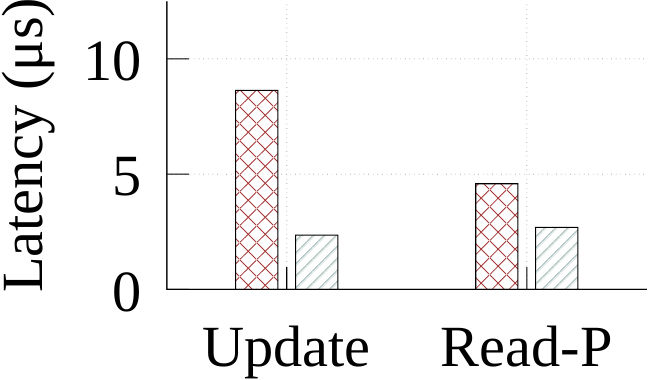}
        \label{fig:selective_access_breakdown}
}
   % \vspace{-10pt}
	\caption{Performance dilemma of adapter layer. The "Read-F/P" indicates reading the full or partial attributes.}
	\label{fig:adapter_layer_analysis}
	\vspace{-10pt}
\end{figure}

% The store was populated with 10 million KV pairs, each 1 KB in size with 10 attributes, following the SQL workload in YCSB.
We conducted experiments on ListDB~\cite{Kim2022listdb} to quantify the impact on performance, following the operation ratio settings in \cite{Cooper10ycsb} to simulate production SQL workloads. The number of attributes in each record is set to 10, with each attribute sized at 100 bytes, consistent with \cite{Cooper10ycsb}. The experimental results (see Figure~\ref{fig:adapter_layer_analysis}) show that when each mapping strategy encounters its unfavorable access pattern (full access for scattered mapping and partial access for consolidated mapping), the latency increases by 1.7$\sim$4.8$\times$ compared to the ideal (i.e., the latency of the other mapping). Thus neither could achieve optimal performance for both full access and partial access. More experimental details are shown in \S\ref{sec:evaluation}.
% Under \emph{consolidated mapping}, while performance is good for \emph{full-row access}, latency is over 3$\times$ higher than \emph{scattered mapping} for \emph{partial-row access} due to \emph{I/O Amplification} and serialization/deserialization costs. Conversely, \emph{scattered mapping} shows low latency for \emph{partial-row access} but incurs over 7$\times$ higher latency for \emph{full-row access} compared to \emph{consolidated mapping}, primarily due to \emph{I/O Splitting} and multiplied indexing overhead. To summarize, due to the above limitations, neither data mapping approach achieves high performance across all request types.

% In selection access workloads, the scenario differs. Consolidated conversion packs data together, requiring reading and rewriting of unrelated attributes. Consequently, the write latency is 3.0$\times$ higher, and the read latency is 3.0$\times$ higher compared to the scattered conversion.

% \begin{figure}[!t]
% \subfigure[Full Access]{
%         \includegraphics[width=0.465\linewidth]{figures/exp_figs/motivation_solutions_full_breakdown.pdf}
%         \label{fig:full_access_breakdown}
% }
% % \hspace{-10pt}
% \subfigure[Selective Access]{
%        \includegraphics[width=0.465\linewidth]{figures/exp_figs/motivation_solutions_partial_breakdown.pdf}
%         \label{fig:selective_access_breakdown}
% }
%    % \vspace{-10pt}
% 	\caption{Performance dilemma of adapter layer.}
% 	\label{fig:adapter_layer_analysis}
% 	\vspace{-10pt}
% \end{figure}

\subsection{Main Idea and Challenges}
\label{subsec:idea}

\begin{figure}[!t]
\centering
\subfigure[Full Read]{
        \includegraphics[width=0.4\linewidth]{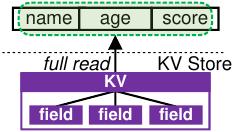}
        \label{fig:focus_efficiency_full}
}
% \hspace{-10pt}
\subfigure[Partial Read]{
       \includegraphics[width=0.4\linewidth]{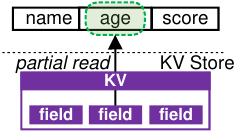}
        \label{fig:focus_efficiency_partial}
}
   % \vspace{-10pt}
	\caption{Hierarchical data management and data mapping.}
	\label{fig:hierarchical_data_management}
	\vspace{-10pt}
\end{figure}

\noindent\textbf{Main idea.}
Recall that the flat and indivisible KV pairs impose inevitable limitations in scenarios with mixed full and partial accesses. To address this complexity, we propose \emph{hierarchical data management}. This approach involves (\romannumeral 1) decomposing KV pairs into finer-grained fields, enabling independent updates and reads on these fields, and (\romannumeral 2) cohesively managing fields belonging to the same KV pair to improve data locality, simplify indexing, and minimize metadata overhead. Therefore, the traditional flat KV pair can be extended into hierarchical ones, each composed of multiple loosely coupled fields.
By mapping a record to such a KV pair, while aligning the schema (i.e., the attributes in the record) with the fields within the KV pair, the upper-layer NewSQL database can selectively read or update desired attributes by operating on the corresponding fields.
For example, we show full read process in Figure~\ref{fig:focus_efficiency_full} and partial read in Figure~\ref{fig:focus_efficiency_partial}. We term such capability \emph{schema-aware access}.

\noindent{\bf Challenges.} Realizing the idea of hierarchical data management in KV stores involves the following design challenges:

\begin{itemize}[leftmargin=*]
\item
\noindent \textbf{Interface design.} The API should maintain the simplicity of and compatibility with original KV interface, while offering essential semantic support for NewSQL databases to ingest schema information, thereby enabling schema-aware access.
\item
\noindent \textbf{Data organization.}
The layout of schema metadata and persistent data should be arranged in a coordinated manner to enable efficient field parsing.
Moreover, for a KV pair in NVM, random updates of fields are more efficient with append writes, yet maintaining field compactness and orderliness favors in-place updates. Balancing them requires careful consideration in data organization.
\item
\noindent \textbf{Indexing and caching.}
Partial access to fields inevitably results in small I/Os on NVM, necessitating an in-memory cache to absorb these operations. However, managing the cache separately from the data index incurs additional probe overhead, even on a cache miss, increasing latency in NVM scenarios. Thus, a global index is needed to provide a unified view of data across all storage tiers. Additionally, the caching policy must be reconsidered to accommodate schema-aware access.
\end{itemize}

 \begin{figure}[h]
	\centering
	\includegraphics[width=1.0\linewidth]{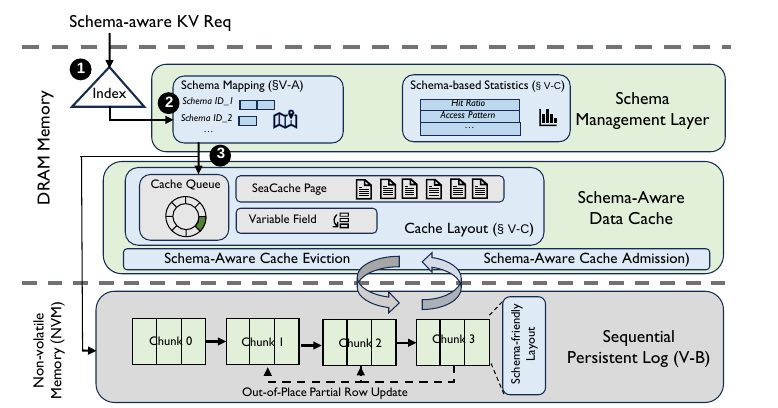}
	\vspace{-20pt}
	\caption{\sysname architecture.}
	 \label{fig:arch}
	% \vspace{-10pt}
\end{figure}

\section{Overview}
\label{sec:overview}

To solve the challenges listed above, we design the \sysname, a log-structured KV store optimized for \underline{f}ine-grained hierarchical data \underline{o}rganization \underline{c}o\underline{u}pled with \underline{s}chema-aware access.
\sysname functions as the stand-alone KV storage engine, offering enhanced schema-oriented access performance. 
Architecturally, as illustrated in Figure~\ref{fig:arch}, \sysname has hierarchical components in DRAM and NVM. 
The first component is the in-memory \textit{schema management layer}, 
responsible for storing schemas and providing quick field location in schema-aware access. 
Data is distributed across DRAM and NVM, with the NVM portion managed by a \textit{persistent log} for durability, and the DRAM portion stored in a \textit{schema-aware cache} to accelerate access. 
These data components are unified by an in-memory \textit{global index} to facilitate querying and updating.

\noindent\textbf{Schema-aware Data Access.} To support the schema-aware access semantically, we extend simple KV APIs by adding schema information when reading or updating data. Table~\ref{tab:API} shows the new APIs of \sysname. 
The first API is \code{create\_schema}, which requires users to provide all field information, including the field name, type, and, most importantly, size.  
In \sysname, a \code{key} is beyond a record that includes the schema identifier, which indicates the schema to which it belongs. and the \code{value} is a set of all data fields (i.e., \{name, age, score\}) in the record. 
In the extended KV API, we introduce an additional parameter, $\mathcal{F}$, in the \code{put}, \code{get}, and \code{scan} functions, which specifies the particular field to access within a KV. 
Thus, \code{$\mathcal{F}$ = \{name, age\}} means the subset contains fields \code{name} and \code{age}.
\sysname APIs allow users to specify the required fields for update, get, and scan operations, thus avoiding unnecessary data access.
To maintain compatibility, if this parameter is left empty, the schema-aware API will automatically default to the original KV interface with full access.

In a nutshell, all data is persisted in NVM within the \PLOG. On a normal read, the memory-resident data index first calculates the starting and ending address for all required fields within the KV pairs using the pre-computed field offset according to the schema. Then, it can access only required fields on NVM without retrieving the irrelevant fields.

We then briefly sketch three main design points that make \sysname efficient and practical. 

\begin{table}[t]
\footnotesize
\renewcommand\arraystretch{1.2}
  \setlength{\tabcolsep}{10pt}
\begin{tabular}{cc} \hline
\cellcolor[HTML]{E1E1E1}{\sysname  API} & \cellcolor[HTML]{E1E1E1}{Description} \\ \hline
 \scriptsize \circled{1} \code{\textbf{create\_schema(name, $\mathcal{F}$)}}   & \scriptsize Create a new schema           \\
 \scriptsize \circled{2} \code{\textbf{put(key, value)} }   & \scriptsize Insert a new record           \\
 \scriptsize \circled{3} \code{\textbf{update(key, value$_\mathcal{F}$)}}   &  \scriptsize  Update the fields ($\mathcal{F}$)           \\
  \scriptsize \circled{4} \code{\textbf{get(key, $\mathcal{F}$)}}   &  \scriptsize  Read the fields ($\mathcal{F}$)        \\
  \scriptsize \circled{5} \code{\textbf{scan(key, $\mathcal{F}$, range)}}  &  \scriptsize Read the fields ($\mathcal{F}$) in the range           \\
  \scriptsize \circled{6} \code{\textbf{del(key)}}  & \scriptsize Delete the corresponding KV          \\   \hline
\end{tabular}
   \vspace{5pt}
\caption{\sysname APIs} 
\label{tab:API}
 \vspace{-20pt}
\end{table}

\noindent\textbf{\S\ref{subsec:schema_friendly_data_layout}: Schema-friendly Layout.} 
Rather than adopting the conventional methods of storing all different fields in a KV pair sequentially (as seen in the \italic{consolidated conversion} approach, see \S\ref{subsec:flat}), or separating all fields into individual KV pairs in an aggressive manner (as observed in the \italic{scattered conversion} KV approach, see \S\ref{subsec:flat}), we introduce a \italic{schema-friendly layout} (see \S\ref{subsec:schema_friendly_data_layout}). This involves categorizing the fields within a given schema into fixed-length and variable-length categories, which are then stored sequentially, employing distinct layouts. The fixed-length data fields maintain a consistent size, ensuring the offset within a KV pair remains constant. For variable-length fields (e.g., strings), it has a fixed-length metadata header, separated from the variable-length content. 

Therefore, a KV pair is divided into two parts: metadata and data. The metadata contains the meta information of the size offset of all fields in the schema, while the data part is stored in \PLOG. 
    
\noindent\textbf{\S\ref{subsec:efficient_update_nvm}:
Sequential Write ahead of In-place Merge}. 
We design to manage data in \PLOG by two sub-logs: \CLOG and \DLOG, based on their completeness.
For the reason that small sequential writes perform more efficiently than random writes, we design a new data update mechanism, which we call \textbf{\italic{swim}} (\textbf{s}equential \textbf{w}rite ahead of \textbf{i}n-place \textbf{m}erge). We let an update directly appended into the persistent log and merge the results into the previous data version asynchronously. Additionally, guided by the tiny writes issues of NVM,  we design a \italic{cacheline aware in-place merge} algorithm (see \S\ref{subsec:efficient_update_nvm}).

\noindent\textbf{\S\ref{subsec:schema_aware_cache}: Schema-aware Caching}.
As the small random reads on tiny fields performs badly on NVM, we bring the schema-aware cache (SeaCache) layer to absorb hot data access. Specifically, SeaCache adopts a schema-aware cache admission and eviction policy to cache hot KV pairs for skewed workloads with a carefully designed data format. 
Compared to the schema-agnostic approach, SeaCache avoids space preemption among different schemas.

\section{Design and Implementation}
In this section, we illustrate our three design points. In a nutshell, 
as shown in Figure~\ref{fig:data_storage}, all data in \sysname is persisted in NVM within the \TLOG (i.e., a two-layered log) in \S\ref{subsec:schema_friendly_data_layout}. We also design efficient delta update mechenaism \italic{swim} in  \S\ref{subsec:efficient_update_nvm} . To speed up skewed workloads, \sysname caches hot data in SACache (to be illustrated in \S\ref{subsec:schema_aware_cache}). 

\label{sec:design}

\subsection{Persistent Data Organization}
\label{subsec:schema_friendly_data_layout}

\begin{figure}[!t]
	\centering
	\includegraphics[width=1.0\linewidth]{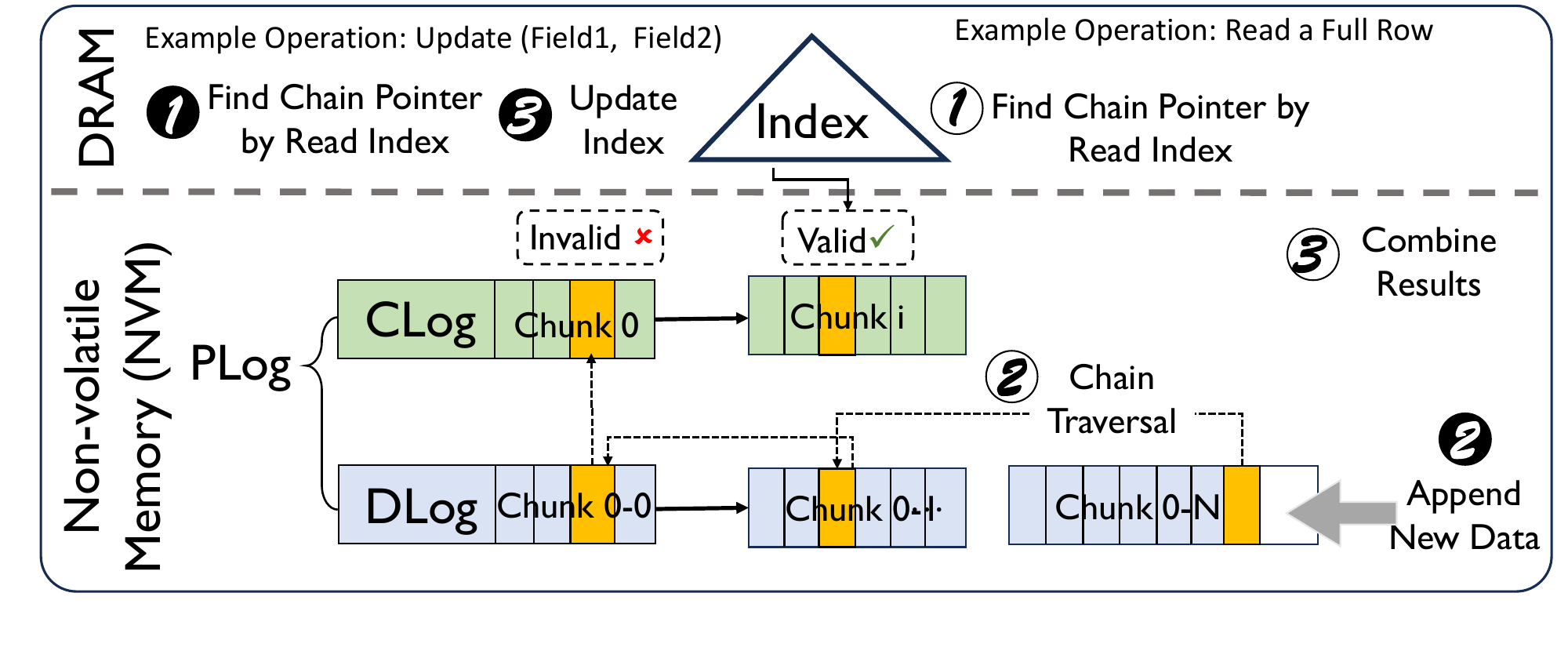}
	\vspace{-20pt}
 \caption{This diagram shows the workflow of partial reads and writes in \sysname when using the out-of-place updates. }
	\label{fig:data_storage}
	\vspace{-10pt}
\end{figure}

To optimize the performance of partial access, we adopted a two-layered persistent log (\PLOG) for the storage under the index layer.
Figure~\ref{fig:data_storage} illustrates our design. We categorize KVs into two types: complete KV pairs and delta updates. Complete KV pairs are stored in the \textit{CLog}, while delta updates are housed within the \textit{DLog}. For management, KV pairs are further grouped into chunks, and each chunk is linked to the nearby chunks for easy management. 
Each DLog is associated with a chunk in the CLog.
When a partial update occurs in any KV within the chunk of the \CLOG, this delta update will be stored in the \DLOG corresponding to the chunk.

% \textcolor{orange}{TODO: unclear definition, delta, entire, monolithic scatter}
To implement partial updates, we decide to update at the granularity of delta updates. 
This approach optimizes storage by only storing incremental changes but also introduces complexities in merging.
To address these challenges, we separate data into two persistent logs: CLog for complete key-value logs and DLog for delta update logs. 
We also designed a minimal overhead merging process \italic{swim} for this data .

\begin{figure}[t] 
	\centering
	\includegraphics[width=1.0\linewidth]{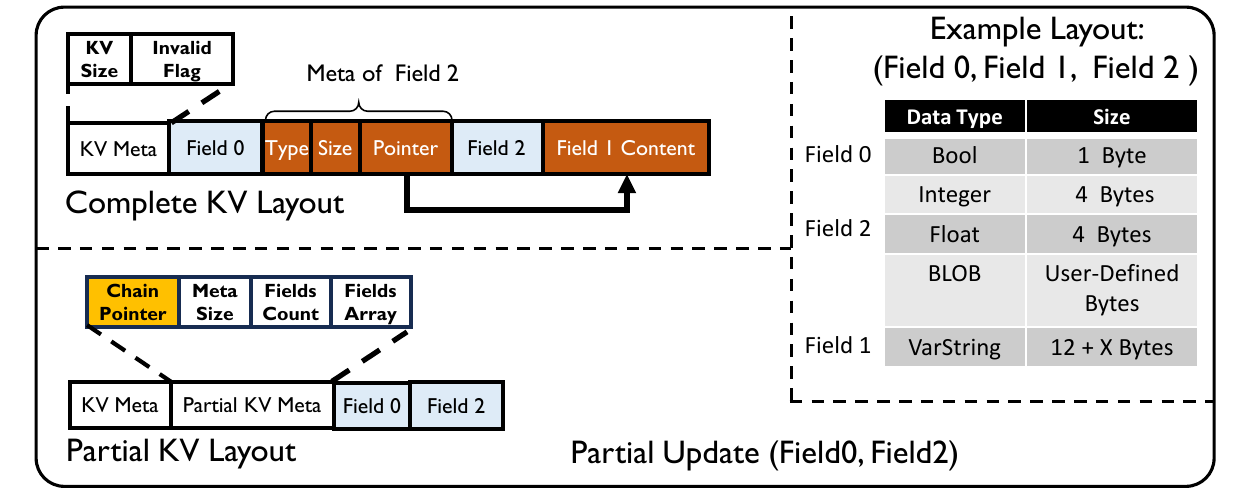}
	\vspace{-20pt}
	\caption{Partial access optimized data layout.}
	\label{fig:layout}
	\vspace{-10pt}
\end{figure}

\noindent\textbf{CLog Layout.} 
\CLOG serves as the primary storage layer in \TLOG, storing the complete KV pairs in an append-only manner.
As shown in Figure~\ref{fig:layout}, a complete KV comprises the KV metadata situated in the KV header. 
The fixed-length content of each field is organized sequentially. 
The KV metadata includes two bytes for KV size and an invalid flag. KV size specifies the whole data size of all fixed- and variable-length content. An invalid flag identifies the existence of a KV in the system.
The fixed-length fields occupy the defined size in KV and store the content directly.

The variable-length fields have two parts: metadata head and content part. 
The metadata head is twelve bytes in total, which consists of two bytes for data types, two bytes for data sizes, and eight bytes for the starting address of the variable-length field. 
As an optimization, we also consider storing those variable-length fields whose size is not over eight bytes directly by reusing the space for the starting address pointer. 
\sysname can easily distinguish pointers or contents by checking the field size (i.e., it indicates a pointer only when the size is bigger than eight bytes).

\noindent\textbf{DLog Layout.} 
\DLOG serves as the second delta update layer and stores all the partial KVs in an append-only manner.
As shown in Figure~\ref{fig:layout}, a partial KV has a partial metadata field in the head that records the field ID, which is 
\code{FieldsCount} and \code{FieldArray} and its corresponding size \code{MetaSize}. \code{ChainPointer} records the address of the preceding entry. Other fields are stored in the order of \code{FieldArray}.

\noindent\textbf{Full Update Process.}  
The full update includes three steps: 1. Write the data to CLOG and obtain the corresponding address on NVM. 2. Construct the corresponding index entry, including the key and its corresponding address. 3. Write this index entry into the index.

\noindent\textbf{Full Read Process.}
The full read is the reverse process of a full update. First, we query the index to get the location of the record in NVM, and then we fully read the record from NVM. 
A more complex scenario involves the record containing multiple subsequent updates, which is detailed in the following section.

\subsection{Efficient Data Update on NVM (\textit{swim})}
\label{subsec:efficient_update_nvm}

We design an efficient data update mechanism,
which we call \textbf{\italic{swim}} (\textbf{s}equential \textbf{w}rite ahead of \textbf{i}n-place \textbf{m}erge).
It contains two steps: instant out-of-place writes and delayed merge.

\noindent\textbf{Out-of-place Update via Sequential Appends.} 
In NVM, sequential writes perform better than random writes (\S\ref{subsec:nvm_perf_characteristic}).
For better update performance, \sysname conducts inserts and updates out-of-place by leveraging a chain-based storage model within the storage layer. 
In our design, a complete row serves as the head, while the remaining partial updates assemble the fields requiring modification into partial rows. These partial rows, in turn, reference the most recent update, thereby creating a linked data chain. To optimize index memory usage, only the address of the last row is stored in the index.

\noindent\textbf{Partial Update Process.}
For the partial row layout, it has an additional meta field that records the field ID requiring an update, which is 
(\code{FieldsCount} and \code{FieldArray}), its corresponding size
(\code{MetaSize}), and the record address (\code{PrevRow}) preceding this
entry. Other fields are stored in the order of \code{FieldArray}.
When addressing the issue of potential data loss during concurrent 
updates within a single-chain structure, the update process becomes a little more complicated.
(i) Query the index for the key to get the corresponding address $r_0$.
(ii) Write data with \code{PrevAddr} set as $r_0$ to \PLOG, then get the new address $r_1$.
(iii) Use a $CAS(r_0,r_1)$ operation to update the index with the new
address atomically. If unsuccessful, update \code{PrevAddr} in \PLOG and repeat until success. This
ensures data consistency without the need for locking mechanisms. 

In addition, a mechanism akin to a restore point is implemented to safeguard
read performance. This mechanism keeps track of the count of consecutive partial
updates. Once this count exceeds a threshold, typically set at 5, it initiates a
full-value write operation. This involves reading all the data and subsequently
rewriting it, effectively reconstructing a complete SKV record. Through this
approach, the system ensures that read and write performance balance, even in
the presence of frequent partial updates, especially in mixed read and write
workloads.

\begin{algorithm}
\footnotesize
    \caption{Cacheline aware Flush}
    \label{alg:cacheline_aware_flush}
\SetKwInOut{Input}{Input}\SetKwInOut{Output}{Output}\SetKwFunction{ORC}{ORC}\SetKwFunction{min}{min}
    \Input{ Updates need to merge $mlist$;}
    % \Output{ Whether write request $r_i$ induces  \emph{stable} traffic or \emph{burst} traffic}
    $flush\_point \leftarrow \code{align}(\text{mlist}[0].\text{addr})$; \textcolor[RGB]{79,125,128}{\tcp{ align returns a cache line aligned address}}

    \For{$\text{m} \textbf{ in } \text{mlist}$}{
        $tmp\_addr \gets \code{align}(\text{m.addr})$;\\
        \textcolor[RGB]{79,125,128}{\tcc{Walk in a request at the granularity of cacheline}}
        \While{$tmp\_addr < \text{m.addr} + \text{m.size}$}
        { 
            \If{$tmp\_addr > \text{flush\_point}$}
            {
                 $\code{flush}(\text{flush\_point})$;\\
                 $flush\_point \gets tmp\_addr$;  
            } 
             $tmp\_addr \gets tmp\_addr + 64$;  \textcolor[RGB]{79,125,128}{\tcp{move to next cacheline}}
        }
    }
    $\code{flush}(\text{flush\_point})$; \textcolor[RGB]{79,125,128}{\tcp{don't forget the last one}}
\end{algorithm}

\noindent\textbf{Cacheline Aware In-place Merge for Fixed-length Fields.} Although the
append-only \PLOG ensures the data integrity, it sacrifices the read efficiency
for the long update chain. However, a noteworthy observation arises when updating
fixed-length fields. After appending the new partial write to \PLOG, updating the corresponding 
field in place in the original full value is safe, as the index
records the new version in the subsequent chain. 
This process is referred to as \textit{in-place merge} and is executed asynchronously,
alleviating it from the critical path. 
This strategic implementation empowers the system to achieve enhanced read
performance while scrupulously preserving the complete value of its original
location.

Tiny writes may lead to repeated flush issues when updating a series of specific fields in NVM in the former reasearch~\cite{Chen2020flatstore}. This necessitates special consideration. We designed a 
\textit{cacheline aware flush} mechanism based on the position layout of the updated field in NVM.

The detailed program is listed in Algorithm~\ref{alg:cacheline_aware_flush}, We first calculate the offsets of all the fields requiring merge, sort them in memory, and flush them one by one into NVM. By managing a \textit{flush\_point} variable, each update assesses the gap between the current update's starting address and the flush point. If this gap is larger than a cache line, a flush operation is initiated from the flush point to the ending address of the current update. Subsequently, the flush point is updated as the ending address of this update. 

\noindent\textbf{Partial Read Process.}
When reading specific fields within
the update chain, the process involves traversing from the tail
to the head. Upon collecting all desired fields in memory,
the partial read can terminate early. For fixed-length fields,
\sysname can bypass the need to consider the KV size in KV
metadata and directly compute the ending address utilizing
the KV layout. When dealing with variable fields, \sysname
must first read the metadata head and subsequently retrieve
the corresponding content accordingly.

\subsection{Schema-aware Caching (SeaCache)}
\label{subsec:schema_aware_cache}
Although during the merge period, if a read occurs, the worst case is that the entire update chain needs to be traversed to find the required data, so we design the \PBRB which can absorb the updates of hot KV pairs in memory, and reduce the impact of read performance degradation during this period. 
Schema-oriented workloads include considerable partial-row access (i.e., massive small reads), which may not be efficient enough in NVM  (\S\ref{subsec:nvm_perf_characteristic}). 
However, the conventional list-based LRU cache structure incurs a high cache miss overhead, which is unacceptable with fast devices~\cite{Qiu23forzencache} (e.g., NVM).
We design SeaCache to handle both hot and small reads on skewed workloads.

\subsubsection{Data Organization}
\label{subsec:data_organization}

\begin{figure*}[!t]
	\centering
	\includegraphics[width=0.9\linewidth]{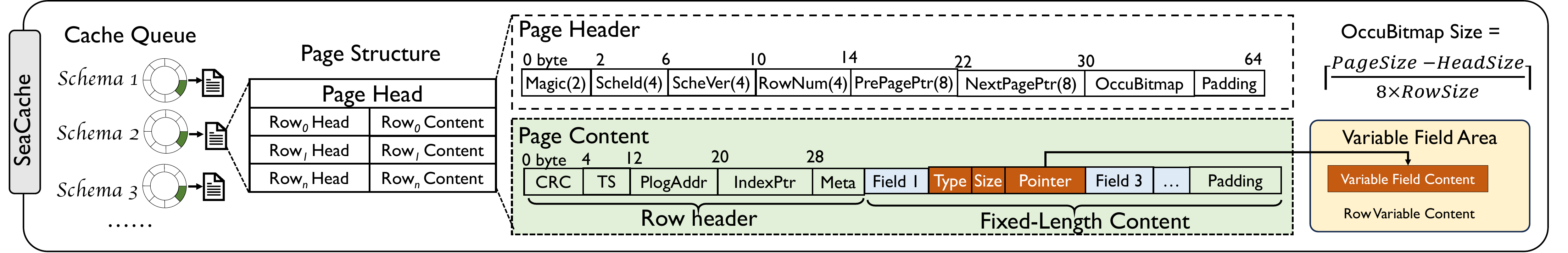}
	\vspace{-5pt}
	\caption{The data structure of SeaCache.}
	\label{fig:page_and_row_details}
	% \vspace{-10pt}
\end{figure*}

As shown in Figure~\ref{fig:arch}, instead of diverging from the conventional cache architecture situated on top of the storage layer, SeaCache is integrated with the index and uses the index to pinpoint the cached data directly. As such, \sysname can find out whether a KV is cached directly from the index pointer, eliminating cache miss fundamentally. We adopt such a design because the cache miss overhead can be unacceptable in the context of NVM~\cite{Qiu23forzencache}. 
For the cache structure, \PBRB uses page lists to store multiple rows. The allocation and reclamation of the cache space are performed at the granularity of memory pages for low metadata overhead. 

\noindent\textbf{Page Structure.} We illustrate the page structure of \PBRB 
in Figure~\ref{fig:page_and_row_details}. Each page is divided into the data section (i.e., rows) and the metadata section (i.e., page header). The data section is divided into a series of fixed-size slots. One slot stores one row using the data format illustrated in \S\ref{subsec:schema_friendly_data_layout}, as well as additional metadata for cache that we illustrated later.

The page header consists of the schema ID of the page, the schema version of the page, the number of cached rows, pointers to the previous and next page, and an occupancy bitmap denoting the usage of empty slots. 
When caching a new row, \PBRB increases the number of rows and sets the corresponding bit of the occupied bitmap. Pages are allocated and incorporated into the page pool when \sysname is initialized, minimizing memory allocation overhead at run-time. When a new page is used, we fetch it from the page pool and insert it into the page list.

\noindent\textbf{Row Structure.} Compared to the data format used in \PLOG, the row structure in SeaCache additionally stores checksum for integrity, the timestamp of the last access for cache coherence, the PLogAddr for the row address in NVM, and the IndexPtr for fast cache eviction (\S\ref{subsec:cache_policy}). Moreover, SeaCache segregates variable-length fields into a designated area to mitigate potential fragmentation issues.

\subsubsection{Cache Policy}
\label{subsec:cache_policy}
Since the access patterns of different schemas may vary greatly, existing schema-agnostic cache policies can easily lead to sub-optimal performance. For instance, suppose there are two schemas: $s_1$ and $s_2$. $s_1$ has a skewed access pattern, while $s_2$ exhibits a more uniform access pattern and has much more concurrency than $s_1$.

When using LRU for cache policy, KVs from $s_1$ and $s_2$ are grouped into the same LRU list. Thus, each KV request for both $s_1$ and $s_2$ leads to a cache admission and eviction. Consequently, a scenario can easily arise in which the rows from $s_1$ are consistently evicted due to a large series data access in $s_2$. 
However, as $s_2$ has a uniform access pattern, the cache hit ratio for the afterward data access can be low. Such a problem is known as a typical cache congestion issue~\cite{Yang2021segcache}, resulting in bad cache performance. 

\noindent\textbf{Admission Policy.}
% When caching SKVs into \PBRB, the most critical issue is to decide which SKVs can be regarded as hot and cached into \PBRB, that is, the design of the admission policy. Before introducing the detailed design, let us first
% \italic{Hit ratio} is one of the most important performance metrics for evaluating cache policies.
In our paper, we use \italic{schema hit ratio} as a metric that captures how many rows can be found in \PBRB given a specific schema. 
% An unselected cache admission will bring extra latency and waste the variable
% cache space. For example, if a large amount of data that will never be accessed
% in the future is cached, they will squeeze other hot data out of the cache and
% take considerable time and resources to expel them, which leads to a low \PBRB cache
% hit ratio and performance jitters. However, 
We determine whether a row should be cached by considering the access pattern of its schema. We calculate the \italic{schema hit ratio} of each schema in the schema management layer (see Figure~\ref{fig:arch}) using the moving average. When initialization (i.e., before the \italic{schema hit ratio} can be calculated), we use the same cache admission policy as LRU (i.e., caching all the latest rows). After that, we set the parameter \bfcode{hit\_threshold} based on the \italic{layout hit ratio} to decide whether a KV should be cached. 
If the row's layout has a higher \italic{layout hit ratio}  than the \bfcode{hit\_threshold}, \PBRB will cache the KV immediately when accessed. Otherwise, \PBRB caches the KV with probability: $P_i=H_i/hit\_threshold$.

\noindent\textbf{Eviction Policy.}
To remove the out-of-date rows efficiently, \PBRB incorporates a schema-aware
cache eviction policy. The key to this policy lies in a \textit{Scheme-aware Lifetime Adjustment}. \PBRB makes the adjustment based on the \textit{schema hit ratio} and data usage within the page list. We assume the schema with higher \textit{schema hit ratio} is more likely to be read again soon. Lifetime is adjusted as follows, where $N$ denotes the number of consecutive failed eviction attempts, $H_{i}$ represents the \textit{schema hit ratio} of $schema _i$, $RO_i$ means the row occupancy of the page list of $schema_i$, and $RW_i$ is a priority parameter.

\vspace{-5pt}
\begin{equation*}
	\label{eq:lifetime_calculation}
	% \vspace{-10pt}
	\mbox{Lifetime}_{i} = 2^{-N}\times H_{i}\times(1-RO_i)\times RW_i
	% \vspace{-5pt}
\end{equation*}

\subsubsection{Cache Workflow} \label{subsubsec:cache_process} \quad

\noindent\textbf{Admission.} One critical issue of cache admission is to find an empty slot in \PBRB. \PBRB adopts a locality-friendly scheme to allocate empty slots sequentially from pages. The execution process includes the following phases: (i) Get an unfull page address via the linked page list. If all pages have been fulfilled, a new page is allocated from the page pool. (ii) \PBRB uses bit-wise operations to find the first zero bit $i$ in the occupancy bitmap, which means the corresponding $slot_i$ that is unused. Then, this slot will be used for a new cache.

\noindent\textbf{Eviction.} The eviction clears out the less frequently accessed rows in \PBRB when the cache space is close to being saturated.
\PBRB considers the rows that have surpassed the \italic{lifetime} threshold 
less likely to be accessed in the near future and are consequently evicted from the cache. To avoid disrupting the foreground service, the eviction is conducted asynchronously, following three steps: (i) start eviction from the schema with the lowest \textit{schema hit ratio} and calculate the survival time for each row in the schema from its last visit time and compare it with the lifetime threshold of the row; (ii) evict all rows whose survival time is larger than the lifetime threshold; (iii) halt the eviction task when the \bfcode{page\_usage} criteria are met. By default, We set the target as 80\%. If the target is not reached, \PBRB updates the lifetime for rows and initiates another round of eviction.

\noindent\textbf{Make the cache workflow NVM-friendly.}
Although \PBRB has a lightweight caching workflow, the process still
requires random access in memory. In our evaluation, cache operation requires approximately $0.6-1.2$${\mu}s$, which significantly impacts the performance.
\PBRB mitigates this issue by removing the cache tasks from the critical read path and pushing them to dedicated background threads. \PBRB uses an asynchronous task queue (i.e., a lock-free circle queue~\cite{Yang16queue}) to record cache requests and uses a separate worker thread to poll the request from the queue.

\section{Evaluation}
\label{sec:evaluation}

%To evaluate the efficiency of schema-aware design, we use the hybrid
%architecture of DRMA+PM and build an embedded KV store engine named \sysname as
%described in \S\ref{subsec:overview}, in which the index structure is selected
%based on the intel thread-safe implementation tbb::concurrent\_map, and we adapt
%the log structure for the PM storage part.

%In this evaluation, we want to answer the following questions about
%our design.
%\begin{itemize}
%	\item How much performance improvement do schema-awareness and \PBRB
%	      achieve compared to other baselines?
%	\item What are the advantages of each design point, and what are the sources
%	of their performance?
%	\item How can we adjust the \PBRB parameter to get the best performance
%	      according to the workload?
%\end{itemize}

\subsection{Experiment Setup}
\label{subsec:experimental_setup}
\noindent\textbf{Implementation.} We implement \sysname from scratch in C++ with $\sim$$5900$ LOC. We use \code{libpmem} for NVM access. For an apple-to-apple comparison, we tune the memory cache size of \sysname to 500\,MB (i.e., the same as the recommended configuration of Pmem-RocksDB and ListDB). We set \bfcode{hit\_threshold} of our cache admission policy as 0.5 and set \bfcode{RW} of our eviction policy according to runtime statistics. 
% In the last experiment, we study the sensitivity of the two parameters. 
To ensure the persistence of written data, we call \code{clfush} and \code{mfence} after each write.

\noindent\textbf{Testbed.} We run all experiments on a two-socket machine equipped with two Intel Xeon Gold 5218R CPUs, 192\,GB DDR4 memory, and 768\,GB Optane DCPMM. 
We create \code{ext4-dax} filesystem on NVM and map the NVM space to the user space using \code{mmap}.

% The machine runs Ubuntu 20.04 LTS with Linux kernel 5.4.80. 
% For NVM access, we utilize the \code{libpmem} library and \code{libpmemobj} library, specifically version 1.11 from Intel.  
% As Pmem-RocksDB and ListDB also use DRAM as a cache, the optimal configuration is setting the cache size to 500MB by integrating the cache tuning guide of RocksDB. So we configure \sysname to use the same amount of memory by setting parameter \bfcode{max\_page\_num} as 12800. For the selective caching strategy, we set  \bfcode{hit\_threshold} as 0.5, implying that the schema with a hit ratio lower than 0.5 will not be cached.

\noindent\textbf{Workload.} We run both the widely studied YCSB~\cite{Cooper10ycsb} and self-created microbenchmarks to evaluate the performance of \sysname. YCSB defines five types of workload (see Table \ref{tab:ycsb_workload}). We set each workload to comprise 10 million operations performed on a randomly populated dataset with 10 million records. The size of each KV is 1KB and consists of 10 fields. ``Read'' operators retrieve a full KV; ``Update'' operators modify a single field inside a KV; ``Insert'' operators insert a new full KV; \textcolor{black}{``Latest reads'' operators read the new inserted full rows; ``Scan'' operators read 100 full KVs; ``Read-modify-write'' operators contain a ``Read'' and a ``Update''.}
% Other parameter settings are kept as default.

\begin{table}[h]
	\centering
	\small
 \vspace{-5pt}
	\begin{tabular}{lcccc}
		\hline
		Workload              & Statistics                         \\ \hline
		A (Update Heavy)      & 50\% updates, 50\% reads           \\
		B (Read Mostly)       & 5\% updates, 95\% reads            \\
		C (Read Only)         & 100\% reads                        \\
		D (Read Latest)       & 5\% inserts, 95\% latest reads     \\
		E (Scan Mostly)       & 5\% inserts, 95\% scans            \\
		F (Read-Modify-Write) & 50\% read-modify-write, 50\% reads \\ \hline
	\end{tabular}
	\vspace{2pt}
	\caption{YCSB workloads.}
	\label{tab:ycsb_workload}
	\vspace{-20pt}
\end{table}

We develop a new microbenchmark to fine-tune the data access patterns. Similar to YCSB, we use 10 million records with an identical number of fields and data sizes.
We extend the four basic operations of YCSB by adding two partial operations, resulting in six basic operations: "Insert," "Read-F (fully)," "Read-P (partially)," "Update," "Scan-F (fully)," and "Scan-P (partially)." 
``Read-P'' reads a field of a row, and ``Scan-P'' scans a field of 100 rows. We consider a skewed access distribution to squeeze the performance of \PBRB. We set the Zipf as 0.99~\cite{Li19elasticbf}.

\noindent\textbf{Baselines.} We evalute three state-of-the-art key-value stores that are heavily optimized persistent memory: Pmem-RocksDB~\cite{Pmem-RocksDB}, and ListDB~\cite{Kim2022listdb}. Among them, Pmem-RocksDB separates the value with \code{libpmemobj} and stores all the keys in LSM-tree; ListDB employs a three-tier skip-list structure, comprising an in-memory caching layer and two persistent layers on PM.

% Given that these three baselines inherently lack schema design and treat values as strings, 

We incorporate two existing mapping approaches (see \S\ref{subsec:flat})) to support partial access on the three KV stores (i.e., consolidated  KV and scattered KV ). We name the delivered systems using suffixes. For example, ListDB (Con) means using the consolidated  KV approach on ListDB, and ListDB (Sct) means using the scattered KV mapping approach.

%The last is
%the setting of $RW$ during the GC process. We choose the best setting in the
%evaluation part\S\ref{subsec:tunable_parameters} according to the hit ratio. We
%will discuss the selection of this parameter in the
%\S\ref{subsec:tunable_parameters}.

\begin{figure}[!t]
	\centering
	\includegraphics[width=1.0\linewidth]{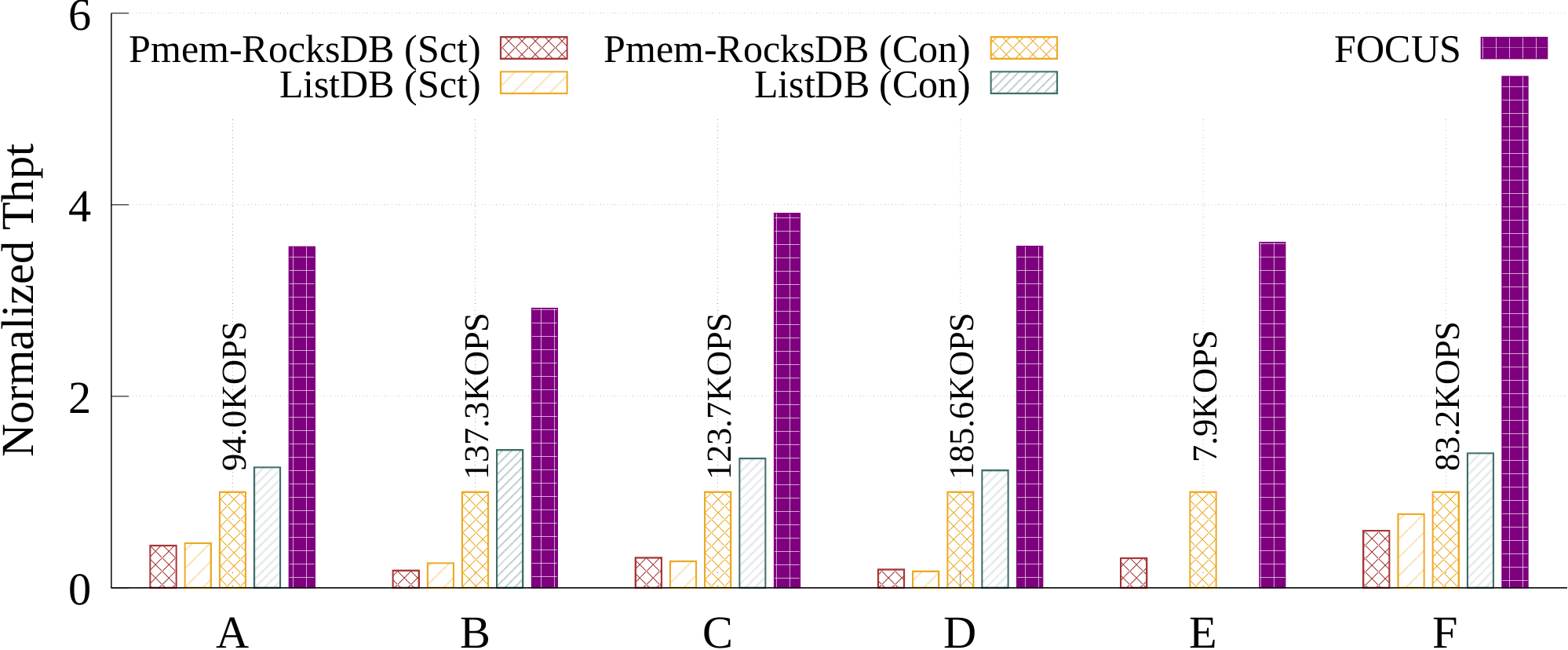}
	\vspace{-10pt}
	\caption{YCSB performance.}
	\label{fig:ycsb_peformance}
	\vspace{-10pt}
\end{figure}

\setlength{\belowcaptionskip}{2pt}
\setlength{\intextsep}{-2pt}

\begin{figure}[!t]
\centering
\includegraphics[width=0.95\linewidth]{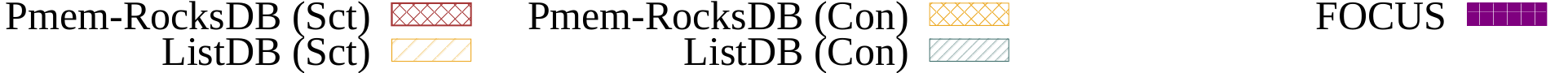}

\subfigure[Full access performance]{
		\centering
     \includegraphics[width=0.465\linewidth]{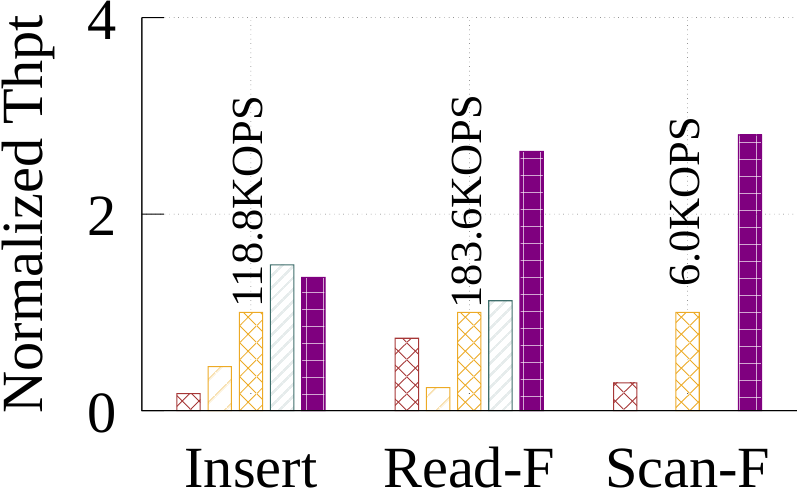}
	 \label{fig:full_access_throughput}
}
\subfigure[Full access latency]{
     \includegraphics[width=0.465\linewidth]{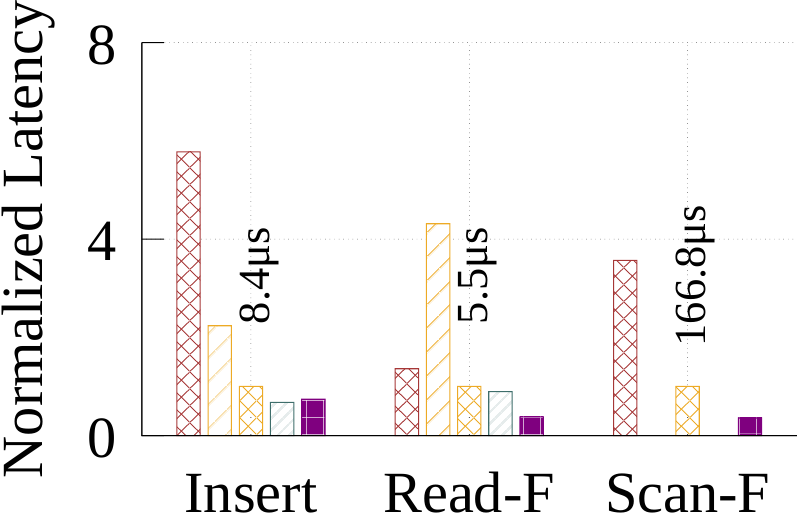}
	 \label{fig:full_access_lat}

}

\vspace{-5pt}
\caption{Performance with full access.} 
\label{fig:microbench_zipfian_schema_full_access}
\vspace{-5pt}
\end{figure}

\begin{figure}[!t]
\centering
\includegraphics[width=0.95\linewidth]{figures/exp_figs/basic_operation_legend.pdf}

\subfigure[Partial access performance]{
		\centering
     \includegraphics[width=0.465\linewidth]{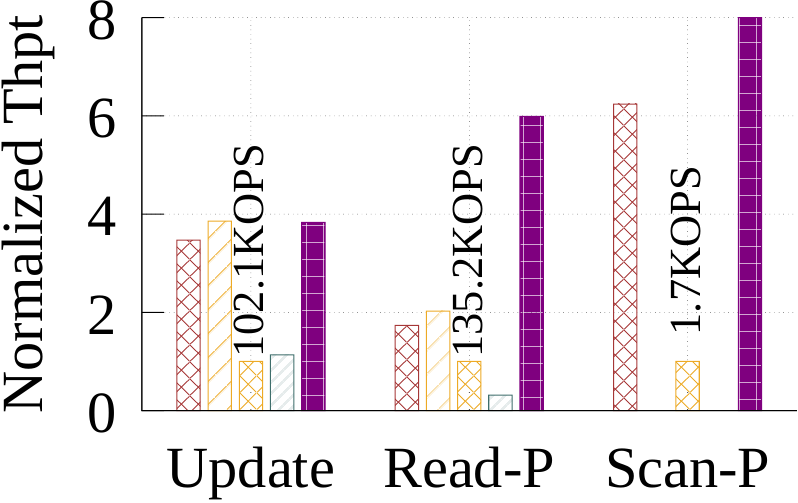}
	 \label{fig:partial_access_throughput}
}
\subfigure[Partial access latency]{
     \includegraphics[width=0.465\linewidth]{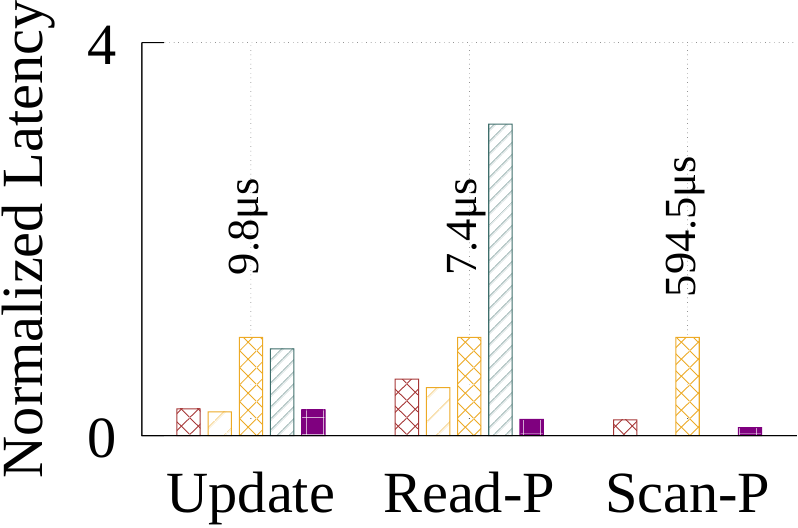}
	 \label{fig:partial_access_lat}
}

\vspace{-5pt}
\caption{Performance with partial access.} 
\label{fig:microbench_zipfian_schema_partial_access}
\vspace{-5pt}
\end{figure}

\subsection{Overall Performance}
\label{subsec:performance_evaluation}

\noindent\textbf{YCSB.} We show the throughput results in Figure~\ref{fig:ycsb_peformance}. For readability, we normalize the results. We use the performance Pmem-RocksDB (RMW) as a unit and provide the absolute number on its bar.
The results show that \sysname achieves 2.2$\times$ throughput than the best of six baselines on workload A and 2.6$\times$ throughput on workload F. 
The significant improvement is primarily due to the joint optimization of delta update and swim merge designs, which are tailored for both read and partial update operations in \sysname. Given that workloads A and F entail simultaneous reads and partial updates, \sysname achieves optimal overall performance.
For read-dominant workloads, \sysname achieves 1.4$\times$ throughput than the best of six baselines on workload B, 1.5$\times$ on workload C, 2.4$\times$ on workload D, and 2.8$\times$ on workload E.
These read-dominated workload advantages reflect the role of \PBRB's advanced decoupled design.
% It can be shown that the column-based encoding approach amplifies the access
% process for full-value access, transforming single data access into access to 10 KV accesses. This results in a significant reduction in execution efficiency.

% In addition to the observations in the microbenchmark, we also find that, in mixed operations that favor both partial updates and full reads (e.g., workload A), the out-of-place updates characteristics of LSM-tree may additionally degrade the performance on reads. In contrast, with \textit{swim} (\S\ref{subsec:efficient_update_nvm}), \sysname mitigates the read amplification without incurring overhead on writes.

% We evaluate the throughput and latency (both average latency and P99 tail latency) for all operations. 
% We consider both skewed
% workload and uniform workload. As they show similar conclusions, we present the details of the results under the skewed workload only in the interest of space.
\noindent\textbf{Microbenchmark.}
We present the throughput and average latency of full and partial KV access in Figure~\ref{fig:microbench_zipfian_schema_full_access}. 
The results are normalized for ease of interpretation.

For full KV access, \sysname achieves advanced performance in ``Read-F'', and ``Scan-F'' than other baselines (i.e., at least 2.2 $\times$ for reads and at least 1.9$\times$ for scan). This is because of the decoupled cache design of \PBRB, which efficiently caches hot data and removes the management overhead of the critical path. For ``insert'', our \sysname also does not lag behind other baselines that use memory for buffering.

For ``Update'', ``Read-P'', and ``Scan-P'', as shown in Figure~\ref{fig:partial_access_throughput} and ~\ref{fig:partial_access_lat}, \sysname outperforms all baselines that use the consolidated  approach. Compared to Pmem-RocksDB, and ListDB, \sysname reduces the average read latency by 56.1\% to 76.7\% and reduces the average update latency by 74.5\% to 82.1\%. This is because leveraging the KV layout definition avoids unnecessary data access; only the required information is read or modified.  

Compared to the scattered approach, \sysname doesn't demonstrate a significant advantage in update operations. However, \sysname achieves much better performance for reads than the scattered approach since the three baselines can easily be trapped in read amplifications.

\begin{figure}[!t]
\centering
\subfigure[Swim impact]{
    \includegraphics[height=0.325\linewidth]{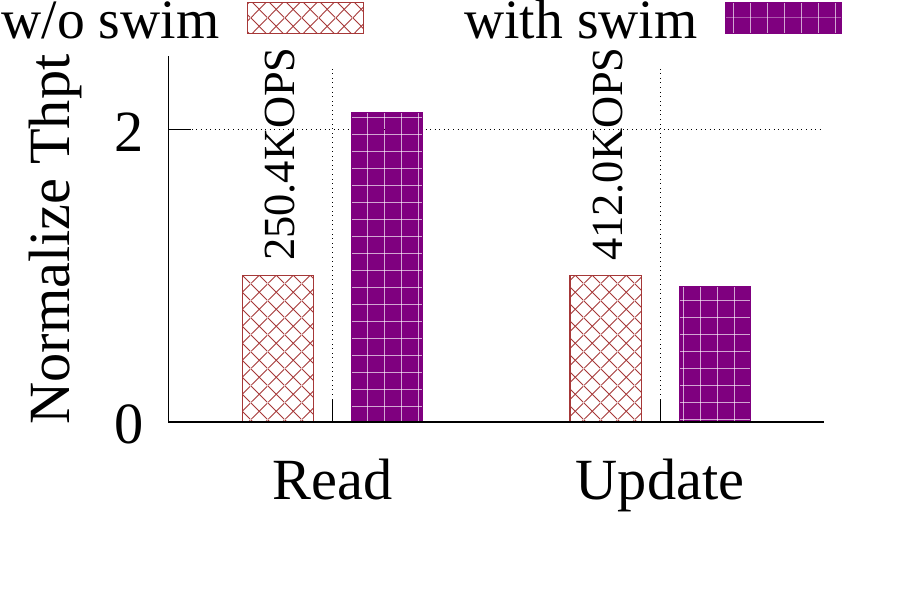}
    \label{fig:analysis_schema_awareness_overall}
}
\subfigure[Cacheline-aware merge impact]{
    \includegraphics[height=0.325\linewidth]{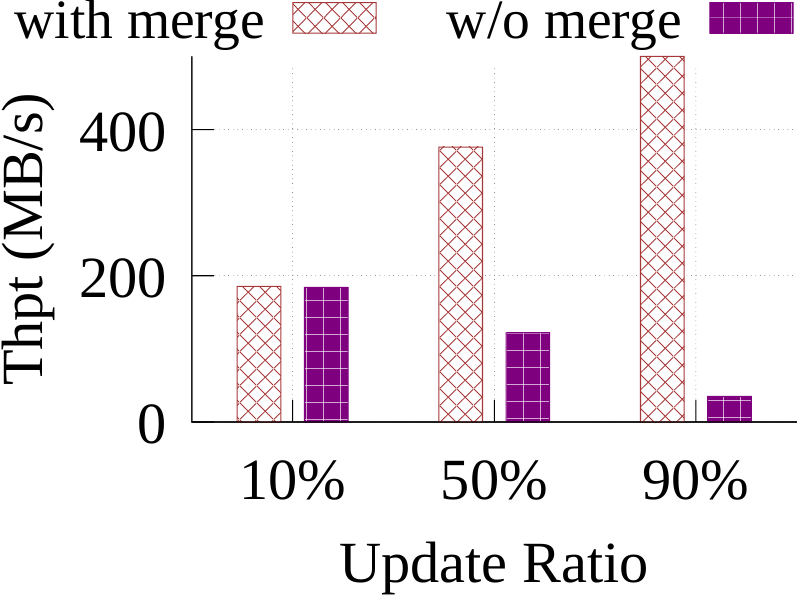}
    \label{fig:analysis_cacheline_aware_flush}
}
\vspace{-5pt}
\caption{Performance analysis of update mechanism.}
\label{fig:analysis_schema_awareness}
\vspace{-5pt}
\end{figure}

\begin{figure}[!t]
\centering
\subfigure[Overall impact]{
    \includegraphics[width=0.465\linewidth]{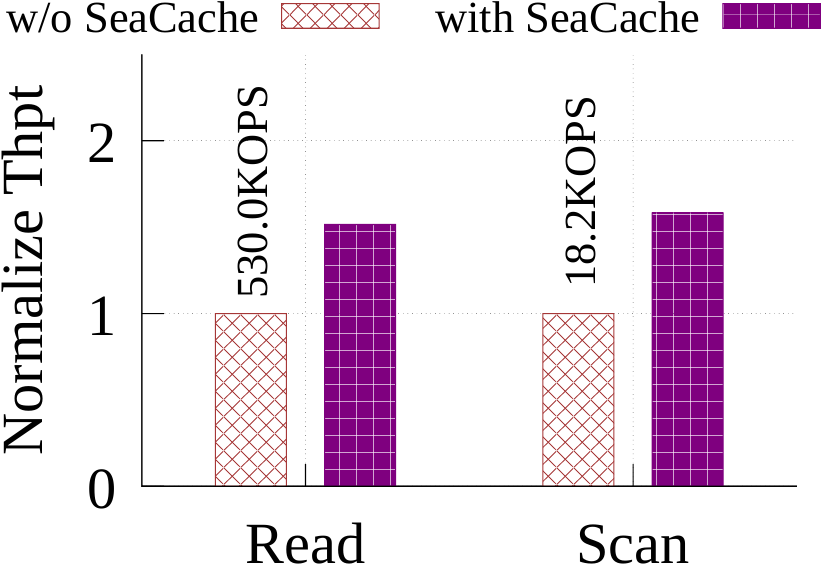}
    \label{fig:analysis_pbrb_overall}
}
\subfigure[Performance breakdown]{
    \includegraphics[width=0.465\linewidth]{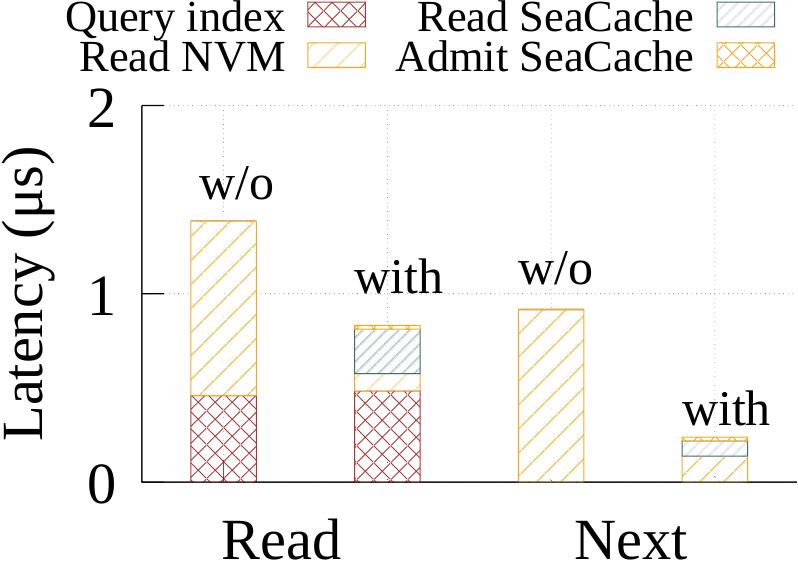}
    \label{fig:analysis_pbrb_breakdown}
}
\vspace{-10pt}
\caption{Effectiveness analysis of \PBRB.}
\label{fig:analysis_pbrb}
\vspace{-10pt}
\end{figure}

\subsection{Performance Breakdown}
\label{subsec:analysis_performance}

In this experiment, we individually study the effectiveness of the designs in \sysname by disabling critical components.

\noindent\textbf{Effectiveness of Swim Mechanism.}{ We first analyze the performance benefits of the \italic{swim}. To do so, we conduct experiments with the \PBRB disabled. Our tests involve reading an entire KV pair and updating an arbitrary KV pair, akin to the YCSB-A benchmark. 
Figure~\ref{fig:analysis_schema_awareness_overall} shows that \sysname improves the read performance by 112.0\% with only 8.3\% update performance sacrifice.  We can conclude that \italic{swim} can greatly improves read performance and only incurs a small update cost.

\noindent\textbf{Effectiveness of Cacheline-aware Merge.}
To investigate the impact of the cacheline aware flush on tiny updates, we configure our KV size with 8 fields, each 32B in size, so the KV size is 256B (aligned with the internal write size of NVM).  We set up different update ratios (from 1/8 to 7/8).

Figure~\ref{fig:analysis_cacheline_aware_flush} illustrates that when the update ratio is low (e.g., 1/8), the impact of cacheline-aware merge is minimal. This is because partial updates only access one field within a 256B-sized KV, thus the problem of repeated flushes does not arise, and the cacheline-aware merge effectively behaves as a regular update.
However, as the update ratio increases, the performance trends become apparent. The throughput of non-optimized updates degrades rapidly, while the cacheline-aware merge improves performance due to the aggregation of internal buffers on NVM. For instance, at the update ratio of 1/8, the cacheline-aware merge improved performance by 13.6$\times$ compared to non-optimized updates.

\noindent \textbf{Effectiveness of \PBRB.} To evaluate the performance benefits of \PBRB, we examine the latency composition of read and scan operations. Figure~\ref{fig:analysis_pbrb} shows the results. As the scan operation commences with a \code{seek()} operation followed by continuous calls to the \code{next()} interface, we only collect the latency composition of\code{next()}. 
Using \PBRB, \sysname achieves 51.5\% and 54.1\% improvement for read and scan operations, respectively, thanks to \PBRB's ability to cache hot data in memory. This also results in a 74.6\% reduction in read latency when accessing cached data.

% \noindent\textbf{Effectiveness of Admission Policy.} We evaluate the effectiveness of the cache admission policy by considering multiple coexisted schemas with different access patterns. Specifically, different access patterns result in different \code{schema hit ratios}. Therefore, we configure five schemas (Schema I-V) with different hit ratios of (i.e., 0.1, 0.3, 0.5, 0.7, and 0.9) for this experiment.  We employ five clients to access the system concurrently, and each client accesses a different schema. The read latency of each schema is measured separately.
% Figure~\ref{fig:multiple_schemas_latency} shows the latency results for both read and update operations accessing partial rows. We find that the latency of the most skewed schema (i.e., Schema V) is 42.3\% to the least skewed one (i.e., Schema I) for the reads and 66.7\% for the updates.
% For the update workload, the performance difference between the most and least skewed schemas becomes smaller: Schema V has 66.7\% of the latency of Schema I. 

% From the results, we conclude that due to the schema-aware cache admission policy in \PBRB, \sysname allows the schema with a higher hit ratio to better use the cache space and thus achieve a better overall performance.

\noindent\textbf{Effectiveness of Eviction Policy.} To study the effectiveness of our eviction policy, we preheat \PBRB before the experiment. We then simulate the scene where the hot spot moves. Figure~\ref{fig:analysis_gc} shows the overall read and scan performance after hot spot moves. \sysname improves the throughput by 49.6\% and 49.1\%, respectively. 

Through real-time monitoring of the cache hit ratio, we find that \PBRB only needs 0.5s to clear enough space to make the hit ratio reach over 0.6, and the hit rate reaches stability after 2.0s. Finally, the hit ratio is almost the same as the workload that has not been polluted. It demonstrates that our lifetime-based eviction policy can effectively identify outdated data.

% \noindent\textbf{Effectiveness of asynchronous caching.} In the previous experiment, we studied the end-to-end performance improvement of \PBRB. We conduct an ablation study to study the performance improvement brought by asynchronous caching optimization.
% Figure~\ref{fig:analysis_async} shows the performance of synchronous and asynchronous caching strategies after a cache miss occurs. 

% On the whole, the asynchronous strategy can improve performance by 11.9\% and 9.8\% for read and scan. By breaking down these two strategies for a read operation, we can find that the latency of the algorithm for synchronously finding empty positions in cache space is high, accounting for the entire read 40.7\% of the operation. In contrast, the asynchronous caching strategy will not block the foreground service as it transfers the task to the background thread, reducing 38.8\% latency. As such, the asynchronous caching strategy is essential when cache misses occur frequently.

\begin{figure}[!t]
\centering
 \subfigure[Overall impact]{
 \includegraphics[width=0.465\linewidth]{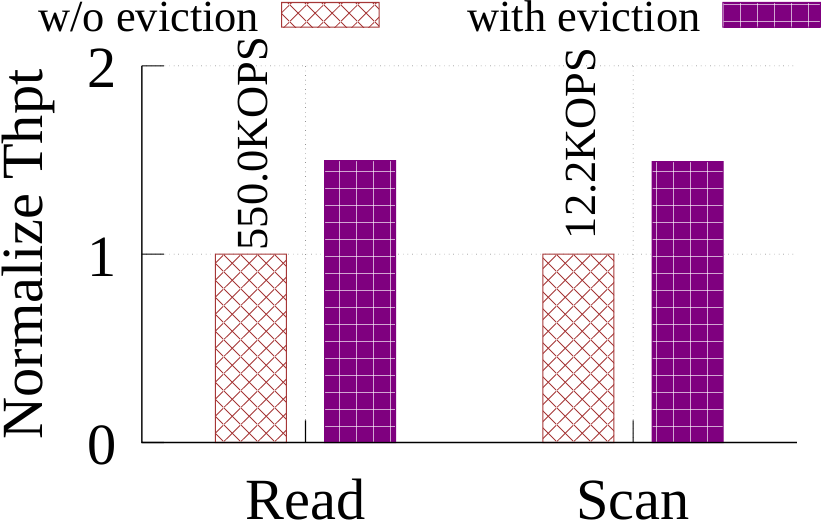}
 \label{fig:analysis_gc_overall}
 }
 \subfigure[Hit ratio over time]{
 \includegraphics[width=0.465\linewidth]{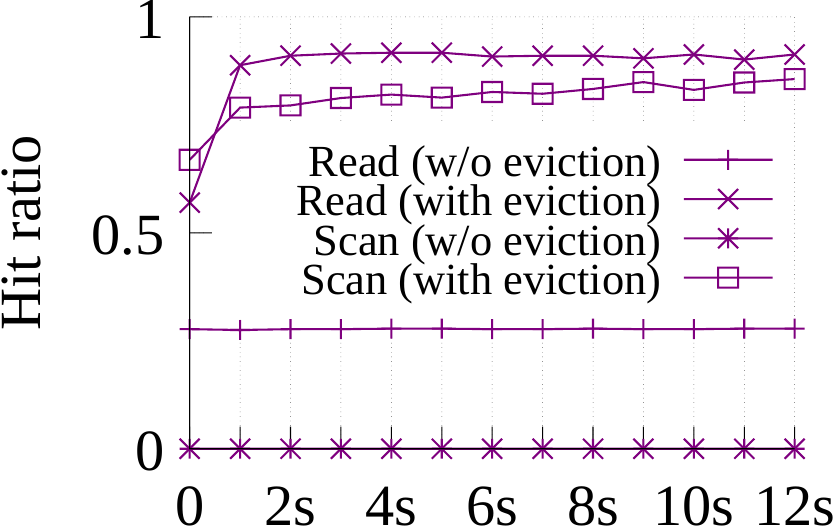}
  \label{fig:analysis_gc_hitratio}
 }
\vspace{-10pt}
\caption{Effectiveness analysis of eviction policy.}
\label{fig:analysis_gc}
\vspace{-10pt}
\end{figure}

\subsection{Parameter Analysis}
\label{subsec:analysis_parameter}
We study the performance of \sysname under diverse workloads and configurations to show its applicability.

% \begin{figure}[!t]
% \centering
% \subfigure[Variable field impact]{
%     \includegraphics[width=0.47\linewidth]{figures/exp_figs/analysis_variable_perf.pdf}
%     \label{fig:analysis_variable_field_overall}
% }
% \subfigure[Storage size]{
%     \includegraphics[width=0.47\linewidth]{figures/exp_figs/anslysis_variable_space.pdf}
%     \label{fig:analysis_variable_field_storage_size}
% }
% \vspace{-10pt}
% \caption{Impact of variable field.}
% \label{fig:analysis_variable_field}
% % \vspace{-10pt}
% \end{figure}

\begin{figure}[!t]
\centering
\subfigure[Multi-thread impact]{
\includegraphics[width=0.45\linewidth]{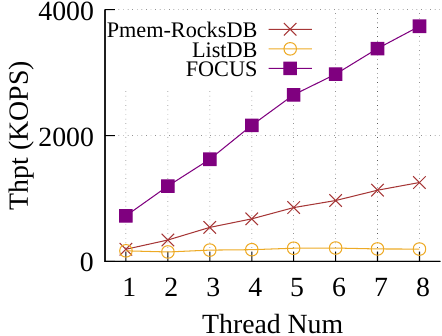}
\label{fig:thread_value_size_analysis_multi_thread}
}
\subfigure[Value size impact]{
\includegraphics[width=0.45\linewidth]{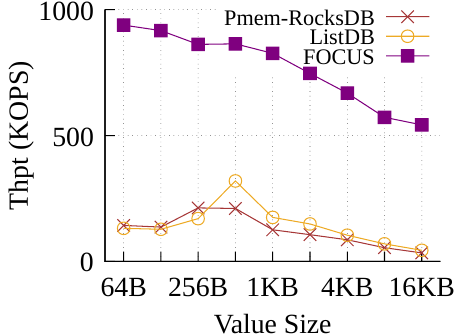}
\label{figure:thread_value_size_analysis_value_size}
}
\vspace{-5pt}
\caption{Impact of multiple threads and value size.}
\label{fig:thread_value_size_analysis}
\vspace{-5pt}
\end{figure}

\noindent\textbf{Impact of Value Size.} To study the impact of value size, we present the performance of \sysname with value sizes varying from 64 Bytes to 16 KB in Figure~\ref{figure:thread_value_size_analysis_value_size}. Since the access granularity of NVM is 256Bytes, any access smaller than this size will cause read amplification.
However, due to the presence of \PBRB, \sysname has a much lower rate of performance drop. The high bandwidth capability of DRAM in transferring small-size values emphasizes the advantages of \PBRB. Even when the value size is 16 KB, the benefits of \sysname reach $7.1\times$. The results demonstrate that \PBRB exhibits notable performance advantages across various value sizes.

\noindent \textbf{Impact of Multi-threads.} To evaluate the scalability in terms of multi-threads, we show the performance curve of \sysname as the number of threads grows (see Figure~\ref{fig:thread_value_size_analysis_multi_thread}). Overall, the throughput almost increases linearly with the number of threads. Compared to the best baseline (i.e., PMem-RocksDB), \sysname's throughput is $3.9-3.7\times$ better. 
Unlike \sysname and PMem-RocksDB, ListDB primarily focuses on write optimization. The result highlights \sysname's ability to efficiently leverage CPU and memory resources, leading to a substantial increase in performance.

\noindent{\bf Sensitivity of Other Tunable Parameters.}
To analyze the sensitivity of parameters, we first study the read throughput of \sysname using YCSB-C by setting different \bfcode{hit\_threshold}. Figure~\ref{fig:parameter_analysis_hit_ratio}
shows the results. Recall that \bfcode{hit\_threshold} is crucial in determining whether a KV should be cached to the \PBRB. 
The curves show the performance with and without the \PBRB intersect at a hit ratio of 0.5. As such, caching data to the \PBRB is only beneficial when the hit ratio
exceeds 0.5. Thus, we set the default value for the \bfcode{hit\_threshold} parameter as 0.5.

We also study the sensitivity of the parameter $RW$, which determines the lifetime of different hit ratios during eviction. We conduct experiments with varying $RW$ values impacts on performance. Figure~\ref{fig:parameter_analysis_eviction_rw} displays the optimal $RW$ value for each hit ratio.
We find if the hit ratio is high or low, small $RW$ can clean up the out-of-date KV pairs quickly without 
affecting the hot data. When the hit ratio is moderate, it's essential to extend the $RW$ appropriately to ensure retention and prevent accidental deletion.
This optimal configuration table of $RW$ is integrated into the system and is continuously updated based on the hit ratio at run-time.

% \noindent \textbf{CPU and space overhead.} \sysname adopts the hybrid
% architecture that stores the full amount of the data in \PLOG and hot data in
% memory, \PBRB needs a separate thread to complete the asynchronous operation, we
% believe that such cost is acceptable for low latency, other poll-based I/O
% frameworks also use similar implementation (e.g., SPDK\cite{spdk}), and as the
% number of threads increases, this overhead will not increase. \PBRB consumes
% acceptable extra memory space compared with the benefits gained from its design.
% Also we argue that real-world workloads usually has a small amount of hot
% data compared with the overall database. Therefore, \PBRB's selective cache
% strategy and schema-aware eviction policy make better usage of memory space. In
% our test, although cutting down the \bfcode{max\_page\_num} to the original 1/4, the
% throughput can still be maintained at the original 92.3\%.

\begin{figure}[!t]
\centering
\subfigure[Hit ratio threshold]{
\includegraphics[width=0.45\linewidth]{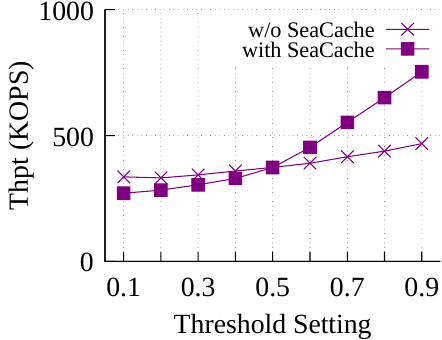}
\label{fig:parameter_analysis_hit_ratio}
}
\subfigure[Eviction retention window]{
\includegraphics[width=0.45\linewidth]{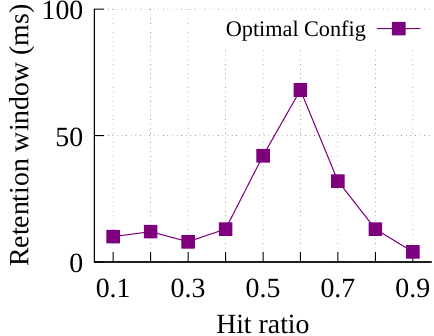}
\label{fig:parameter_analysis_eviction_rw}
}
\vspace{-5pt}
\caption{Sensitivity of tunable parameters in \sysname.}
\label{fig:parameter_analysis}
\vspace{-5pt}
\end{figure}

% \subsection{Efficient Schema Management}
% \label{subsec:schema_awareness}
% In this experiment, we evaluate how efficient the schema management of \sysname is by testing it under different 

\section{Related Works}
\label{sec:related}

\noindent\textbf{NVM-optmized Storage.} KV stores in this category adapt the legacy tiered architecture to the new storage hierarchy. A notable example is NVM-optimized LSM trees, where data is buffered in DRAM and gradually merged into storage. Traditional LSM trees face performance issues like high write amplification and unstable tail latency when using HDD or SSD. NVM can mitigate these issues by leveraging its hardware characteristics.
Key implementations include NoveLSM~\cite{Kannan2018redesigning}, SLM-DB~\cite{Kaiyrakhmet2019slmdb}, MatrixKV~\cite{Yao2020matrixkv}, and ListDB~\cite{Kim2022listdb}. SLM-DB reduces compaction by using a single persistent layer and a B+Tree index. MatrixKV optimizes L0 compaction with a matrix container using NVM. ListDB unifies the write-ahead log and memtable to reduce data movement and employs a two-layer architecture with zipper compaction for efficiency. BonsaiKV~\cite{Cai2023bonsaikv} separates indexing, persistence, and scalability across different hardware layers, addressing NVMM congestion and small write issues with novel techniques.

\noindent\textbf{KV powered applications.}
Many applications convert their specific format of data (e.g., graph structures, table rows, AI vector embeddings) into KV
pairs and then employ existing KV stores to manage their data, e.g., CockroachDB~\cite{Taft20cockroachdb} and TiDB~\cite{Huang2020tidb} use the LSM-tree-based RocksDB~\cite{RocksDB} as their storage engine. These designs follow the row-based storage and require extra designs
to express the schema structure.

% Different from these related works, \sysname targets the efficiency of partial-access-dominated workloads, which is a problem underappreciated by the research community. 
% \input{sections/hierarchical kv management}
% \input{overview}
% \input{design}
% \input{evaluation}
% \input{relatedwork}
\section{Conclusion}
In this paper, we introduced \sysname, a log-structured KV store designed to optimize fine-grained hierarchical data organization and schema-aware access. \sysname addresses the limitations of traditional KV stores, which often rely on flat data mapping approaches that lead to significant I/O amplification and splitting.
Our experimental results demonstrate that \sysname significantly outperforms existing NVM-backed KV stores, such as Pmem-RocksDB and ListDB, under production workload.

%%%%%%% -- PAPER CONTENT ENDS -- %%%%%%%%

%%%%%%%%% -- BIB STYLE AND FILE -- %%%%%%%%
\bibliographystyle{IEEEtranS}
\bibliography{refs}

% Generated by IEEEtranS.bst, version: 1.13 (2008/09/30)
\begin{thebibliography}{10}
\providecommand{\url}[1]{#1}
\csname url@samestyle\endcsname
\providecommand{\newblock}{\relax}
\providecommand{\bibinfo}[2]{#2}
\providecommand{\BIBentrySTDinterwordspacing}{\spaceskip=0pt\relax}
\providecommand{\BIBentryALTinterwordstretchfactor}{4}
\providecommand{\BIBentryALTinterwordspacing}{\spaceskip=\fontdimen2\font plus
\BIBentryALTinterwordstretchfactor\fontdimen3\font minus \fontdimen4\font\relax}
\providecommand{\BIBforeignlanguage}[2]{{%
\expandafter\ifx\csname l@#1\endcsname\relax
\typeout{** WARNING: IEEEtranS.bst: No hyphenation pattern has been}%
\typeout{** loaded for the language `#1'. Using the pattern for}%
\typeout{** the default language instead.}%
\else
\language=\csname l@#1\endcsname
\fi
#2}}
\providecommand{\BIBdecl}{\relax}
\BIBdecl

\bibitem{Aghayev19file}
A.~Aghayev, S.~Weil, M.~Kuchnik, M.~Nelson, G.~R. Ganger, and G.~Amvrosiadis, ``File systems unfit as distributed storage backends: lessons from 10 years of ceph evolution,'' in \emph{Proc. of ACM SOSP}.\hskip 1em plus 0.5em minus 0.4em\relax ACM, Oct. 2019, pp. 353--369.

\bibitem{Akinaga2010resistive}
H.~Akinaga and H.~Shima, ``Resistive random access memory (reram) based on metal oxides,'' \emph{Proceedings of the IEEE}, vol.~98, no.~12, pp. 2237--2251, Dec. 2010.

\bibitem{DynamoDB12}
Amazon, ``Dynamodb homepage,'' \url{https://aws.amazon.com/dynamodb/}, 2012.

\bibitem{Oracle95}
O.~and/or~its affiliates., ``Oracle homepage,'' \url{https://www.oracle.com/database/}, 1993.

\bibitem{Benson2021viper}
L.~Benson, H.~Makait, and T.~Rabl, ``Viper: an efficient hybrid pmem-dram key-value store,'' \emph{Proc. of VLDB Endow.}, vol.~14, no.~9, pp. 1544--1556, May 2021.

\bibitem{Cai2023bonsaikv}
M.~Cai, J.~Shen, Y.~Yuan, Z.~Qu, and B.~Ye, ``Bonsaikv: Towards fast, scalable, and persistent key-value stores with tiered, heterogeneous memory system,'' \emph{Proc. VLDB Endow.}, vol.~17, no.~4, p. 726–739, mar 2024.

\bibitem{caow2020polardb}
C.~Cao, Y.~Liu, Z.~Cheng, N.~Zheng, W.~Li, W.~Wu, L.~Ouyang, P.~Wang, Y.~Wang, R.~Kuan, Z.~Liu, F.~Zhu, and T.~Zhang, ``{{POLARDB} Meets Computational Storage: Efficiently Support Analytical Workloads in Cloud-Native Relational Database},'' in \emph{Proc. of USENIX FAST}, 2020.

\bibitem{Cao22polardb}
W.~Cao, F.~Li, G.~Huang, J.~Lou, J.~Zhao, D.~He, M.~Sun, Y.~Zhang, S.~Wang, X.~Wu, H.~Liao, Z.~Chen, X.~Fang, M.~Chen, C.~Liang, Y.~Luo, H.~Wang, S.~Wang, Z.~Ma, X.~Yang, X.~Peng, Y.~Ruan, Y.~Wang, J.~Zhou, J.~Wang, Q.~Hu, and J.~Kang, ``Polardb-x: An elastic distributed relational database for cloud-native applications,'' in \emph{Proc. of IEEE ICDE}.\hskip 1em plus 0.5em minus 0.4em\relax IEEE, Apr. 2022, pp. 2859--2872.

\bibitem{Chen2020flatstore}
Y.~Chen, Y.~Lu, F.~Yang, Q.~Wang, Y.~Wang, and J.~Shu, ``Flatstore: An efficient log-structured key-value storage engine for persistent memory,'' in \emph{Proc. of ACM ASPLOS}.\hskip 1em plus 0.5em minus 0.4em\relax ACM, Mar. 2020, pp. 1077--1091.

\bibitem{Cooper10ycsb}
B.~F. Cooper, A.~Silberstein, E.~Tam, R.~Ramakrishnan, and R.~Sears, ``{Benchmarking Cloud Serving Systems with YCSB},'' in \emph{Proc. of ACM SoCC}.\hskip 1em plus 0.5em minus 0.4em\relax ACM, Jun. 2010, p. 143–154.

\bibitem{Corbett13spanner}
J.~C. Corbett, J.~Dean, M.~Epstein, A.~Fikes, C.~Frost, J.~J. Furman, S.~Ghemawat, A.~Gubarev, C.~Heiser, P.~Hochschild \emph{et~al.}, ``Spanner: Google’s globally distributed database,'' \emph{{Proc. of ACM TOCS}}, vol.~31, no.~3, pp. 1--22, Aug. 2013.

\bibitem{TPCC}
T.~P.~P. Council., ``{TPC-C} standard specification revision 5.11,'' \url{https://www.tpc.org/tpcc/}, 2010.

\bibitem{Duan2021hardware}
\BIBentryALTinterwordspacing
Z.~Duan, H.~Lu, H.~Liu, X.~Liao, H.~Jin, Y.~Zhang, and S.~Wu, ``Hardware-supported remote persistence for distributed persistent memory,'' in \emph{Proceedings of the International Conference for High Performance Computing, Networking, Storage and Analysis}, ser. SC '21.\hskip 1em plus 0.5em minus 0.4em\relax New York, NY, USA: Association for Computing Machinery, 2021. [Online]. Available: \url{https://doi.org/10.1145/3458817.3476194}
\BIBentrySTDinterwordspacing

\bibitem{Dulloor2016data}
\BIBentryALTinterwordspacing
S.~R. Dulloor, A.~Roy, Z.~Zhao, N.~Sundaram, N.~Satish, R.~Sankaran, J.~Jackson, and K.~Schwan, ``Data tiering in heterogeneous memory systems,'' in \emph{Proceedings of the Eleventh European Conference on Computer Systems}, ser. EuroSys '16.\hskip 1em plus 0.5em minus 0.4em\relax New York, NY, USA: Association for Computing Machinery, 2016. [Online]. Available: \url{https://doi.org/10.1145/2901318.2901344}
\BIBentrySTDinterwordspacing

\bibitem{RocksDB}
Facebook, ``{RocksDB homepage},'' \url{http://rocksdb.org/}, 2013.

\bibitem{Myrocks16}
Facebook, ``Myrocks homepage,'' \url{https://myrocks.io/}, 2016.

\bibitem{HBase}
A.~S. Foundation., ``Hbase homepage,'' \url{https://hbase.apache.org/}, 2008.

\bibitem{Gilmer2018nram}
D.~C. Gilmer, T.~Rueckes, and L.~Cleveland, ``Nram: a disruptive carbon-nanotube resistance-change memory,'' \emph{Nanotechnology}, vol.~29, no.~13, p. 134003, Apr. 2018.

\bibitem{Bigtable15}
Google, ``Bigtable homepage,'' \url{https://cloud.google.com/bigtable/}, 2015.

\bibitem{Gotze2020data}
\BIBentryALTinterwordspacing
P.~G\"{o}tze, A.~K. Tharanatha, and K.-U. Sattler, ``Data structure primitives on persistent memory: an evaluation,'' ser. DaMoN '20.\hskip 1em plus 0.5em minus 0.4em\relax New York, NY, USA: Association for Computing Machinery, 2020. [Online]. Available: \url{https://doi.org/10.1145/3399666.3399900}
\BIBentrySTDinterwordspacing

\bibitem{Hady2017platform}
F.~T. Hady, A.~Foong, B.~Veal, and D.~Williams, ``Platform storage performance with 3d xpoint technology,'' \emph{Proceedings of the IEEE}, vol. 105, no.~9, pp. 1822--1833, Aug. 2017.

\bibitem{Harter4analysis}
T.~Harter, D.~Borthakur, S.~Dong, A.~S. Aiyer, L.~Tang, A.~C. Arpaci-Dusseau, and R.~H. Arpaci-Dusseau, ``{Analysis of HDFS under HBase: A Facebook Messages Case Study},'' in \emph{Proc. of USENIX FAST}.\hskip 1em plus 0.5em minus 0.4em\relax USENIX Association, Feb. 2014, pp. 199--212.

\bibitem{Huang2020tidb}
D.~Huang, Q.~Liu, Q.~Cui, Z.~Fang, X.~Ma, F.~Xu, L.~Shen, L.~Tang, Y.~Zhou, M.~Huang \emph{et~al.}, ``Tidb: a raft-based htap database,'' \emph{Proc. of VLDB Endow.}, vol.~13, no.~12, pp. 3072--3084, Aug. 2020.

\bibitem{Pmem-RocksDB}
Intel., ``{A version of RocksDB that uses persistent memory},'' \url{https://github.com/pmem/Pmem-RocksDB}, 2018.

\bibitem{jiang2020hologres}
X.~Jiang, Y.~Hu, Y.~Xiang, G.~Jiang, X.~Jin, C.~Xia, W.~Jiang, J.~Yu, H.~Wang, Y.~Jiang, J.~Ma, L.~Su, and K.~Zeng, ``{Alibaba Hologres: A Cloud-Native Service for Hybrid Serving/Analytical Processing},'' \emph{Proc. of VLDB Endow.}, vol.~13, no.~12, pp. 3272--3284, 2020.

\bibitem{izraelevitz2019basic}
\BIBentryALTinterwordspacing
I.~Joseph, Y.~Jian, Z.~Lu, K.~Juno, L.~Xiao, M.~Amirsaman, J.~S. Yun, W.~Zixuan, X.~Yi, R.~D. Subramanya, Z.~Jishen, and S.~Steven, ``Basic performance measurements of the intel optane dc persistent memory module,'' 2019. [Online]. Available: \url{https://arxiv.org/abs/1903.05714}
\BIBentrySTDinterwordspacing

\bibitem{Kaiyrakhmet2019slmdb}
O.~Kaiyrakhmet, S.~Lee, B.~Nam, S.~H. Noh, and Y.-r. Choi, ``{SLM-DB}: {Single-Level} {Key-Value} store with persistent memory,'' in \emph{Proc. of USENIX FAST}.\hskip 1em plus 0.5em minus 0.4em\relax USENIX Association, Feb. 2019, pp. 191--205.

\bibitem{Kannan2018redesigning}
S.~Kannan, N.~Bhat, A.~Gavrilovska, A.~Arpaci-Dusseau, and R.~Arpaci-Dusseau, ``Redesigning lsms for nonvolatile memory with novelsm,'' in \emph{Proc. of USENIX ATC}.\hskip 1em plus 0.5em minus 0.4em\relax USENIX Association, Jul. 2018, pp. 993--1005.

\bibitem{kato2007overview}
Y.~Kato, Y.~Kaneko, H.~Tanaka, K.~Kaibara, S.~Koyama, K.~Isogai, T.~Yamada, and Y.~Shimada, ``Overview and future challenge of ferroelectric random access memory technologies,'' \emph{Japanese Journal of Applied Physics}, vol.~46, no.~4S, p. 2157, Apr. 2007.

\bibitem{Kim2005reliability}
K.~Kim and S.~J. Ahn, ``Reliability investigations for manufacturable high density pram,'' in \emph{Proc. of IEEE IRPS}.\hskip 1em plus 0.5em minus 0.4em\relax IEEE, Apr. 2005, pp. 157--162.

\bibitem{Kim2022listdb}
W.~Kim, C.~Park, D.~Kim, H.~Park, Y.~ri~Choi, A.~Sussman, and B.~Nam, ``{ListDB}: Union of {Write-Ahead} logs and persistent {SkipLists} for incremental checkpointing on persistent memory,'' in \emph{Proc. of USENIX OSDI}.\hskip 1em plus 0.5em minus 0.4em\relax USENIX Association, Jul. 2022, pp. 161--177.

\bibitem{Aapo23GraphChi}
A.~Kyrola, G.~Blelloch, and C.~Guestrin, ``{GraphChi: Large-Scale Graph Computation on Just a {PC}},'' in \emph{Proc. of USENIX OSDI}.\hskip 1em plus 0.5em minus 0.4em\relax USENIX Association, Oct. 2012, pp. 31--46.

\bibitem{Li19elasticbf}
Y.~Li, C.~Tian, F.~Guo, C.~Li, and Y.~Xu, ``{Elasticbf: Elastic Bloom Filter with Hotness Awareness for Boosting Read Performance in Large Key-Value Stores},'' in \emph{Proc. of USENIX ATC}.\hskip 1em plus 0.5em minus 0.4em\relax USENIX Association, Feb. 2019, pp. 739--752.

\bibitem{Mao2012cache}
\BIBentryALTinterwordspacing
Y.~Mao, E.~Kohler, and R.~T. Morris, ``Cache craftiness for fast multicore key-value storage,'' in \emph{Proceedings of the 7th ACM European Conference on Computer Systems}, ser. EuroSys '12.\hskip 1em plus 0.5em minus 0.4em\relax New York, NY, USA: Association for Computing Machinery, 2012, p. 183–196. [Online]. Available: \url{https://doi.org/10.1145/2168836.2168855}
\BIBentrySTDinterwordspacing

\bibitem{Mehra2004fast}
P.~Mehra and S.~Fineberg, ``Fast and flexible persistence: the magic potion for fault-tolerance, scalability and performance in online data stores,'' in \emph{Proc. of IEEE IPDPS}.\hskip 1em plus 0.5em minus 0.4em\relax IEEE, Apr. 2004, p. 206.

\bibitem{O1996log}
P.~O’Neil, E.~Cheng, D.~Gawlick, and E.~O’Neil, ``The log-structured merge-tree (lsm-tree),'' \emph{Acta Informatica}, vol.~33, pp. 351--385, 1996.

\bibitem{pilman2017fastscan}
M.~Pilman, K.~Bocksrocker, L.~Braun, R.~Marroquin, and D.~Kossmann, ``{Fast Scans on Key-Value Stores},'' \emph{Proc. of VLDB Endow.}, vol.~10, no.~11, pp. 1526--1537, 2017.

\bibitem{Qiu23forzencache}
Z.~Qiu, J.~Yang, J.~Zhang, C.~Li, X.~Ma, Q.~Chen, M.~Yang, and Y.~Xu, ``Frozenhot cache: Rethinking cache management for modern hardware,'' in \emph{Proc. of ACM EuroSys}.\hskip 1em plus 0.5em minus 0.4em\relax ACM Association, Jul. 2023, p. 557–573.

\bibitem{Raoux2008phase}
S.~Raoux, G.~W. Burr, M.~J. Breitwisch, C.~T. Rettner, Y.-C. Chen, R.~M. Shelby, M.~Salinga, D.~Krebs, S.-H. Chen, H.-L. Lung \emph{et~al.}, ``Phase-change random access memory: A scalable technology,'' \emph{IBM Journal of Research and Development}, vol.~52, no. 4.5, pp. 465--479, Jul. 2008.

\bibitem{ScyllaDB15}
ScyllaDB, ``Scylladb homepage,'' \url{https://www.scylladb.com/}, 2015.

\bibitem{Taft20cockroachdb}
R.~Taft, I.~Sharif, A.~Matei, N.~VanBenschoten, J.~Lewis, T.~Grieger, K.~Niemi, A.~Woods, A.~Birzin, R.~Poss \emph{et~al.}, ``Cockroachdb: The resilient geo-distributed sql database,'' in \emph{Proc. of ACM SIGMOD}.\hskip 1em plus 0.5em minus 0.4em\relax ACM, Jun. 2020, pp. 1493--1509.

\bibitem{Tulapurkar2005spin}
A.~Tulapurkar, Y.~Suzuki, A.~Fukushima, H.~Kubota, H.~Maehara, K.~Tsunekawa, D.~Djayaprawira, N.~Watanabe, and S.~Yuasa, ``Spin-torque diode effect in magnetic tunnel junctions,'' \emph{Nature}, vol. 438, no. 7066, pp. 339--342, Nov. 2005.

\bibitem{Yang16queue}
C.~Yang and J.~Mellor-Crummey, ``A wait-free queue as fast as fetch-and-add,'' in \emph{Proc. of ACM PPoPP}.\hskip 1em plus 0.5em minus 0.4em\relax ACM, Feb. 2016, pp. 1--13.

\bibitem{Yang20empirical}
J.~Yang, J.~Kim, M.~Hoseinzadeh, J.~Izraelevitz, and S.~Swanson, ``An empirical guide to the behavior and use of scalable persistent memory,'' in \emph{Proc. of USENIX FAST}.\hskip 1em plus 0.5em minus 0.4em\relax USENIX Association, Feb. 2020, pp. 169--182.

\bibitem{Yang2021segcache}
J.~Yang, Y.~Yue, and R.~Vinayak, ``Segcache: a memory-efficient and scalable in-memory key-value cache for small objects,'' in \emph{Proc. of USENIX NSDI}.\hskip 1em plus 0.5em minus 0.4em\relax USENIX Association, Apr. 2021, pp. 503--518.

\bibitem{yang2022oceanbase}
Z.~Yang, C.~Yang, F.~Han, M.~Zhuang, B.~Yang, Z.~Yang, X.~Cheng, Y.~Zhao, W.~Shi, H.~Xi \emph{et~al.}, ``Oceanbase: a 707 million tpmc distributed relational database system,'' \emph{Proc. of VLDB Endow.}, vol.~15, no.~12, pp. 3385--3397, 2022.

\bibitem{Yao2020matrixkv}
T.~Yao, Y.~Zhang, J.~Wan, Q.~Cui, L.~Tang, H.~Jiang, C.~Xie, and X.~He, ``Matrixkv: Reducing write stalls and write amplification in lsm-tree based kv stores with matrix container in nvm,'' in \emph{Proc. of USENIX ATC}.\hskip 1em plus 0.5em minus 0.4em\relax USENIX Association, Jul. 2020, pp. 17--31.

\bibitem{YugabyteDB17}
I.~Yugabyte, ``Yugabytedb homepage,'' \url{https://www.yugabyte.com/}, 2017.

\bibitem{Zhang2021chameleondb}
W.~Zhang, X.~Zhao, S.~Jiang, and H.~Jiang, ``Chameleondb: a key-value store for optane persistent memory,'' in \emph{Proc. of ACM EuroSys}.\hskip 1em plus 0.5em minus 0.4em\relax ACM, Jul. 2021, pp. 194--209.

\bibitem{Da15FlashGraph}
D.~Zheng, D.~Mhembere, R.~Burns, J.~Vogelstein, C.~E. Priebe, and A.~S. Szalay, ``{FlashGraph: Processing Billion-Node Graphs on an Array of Commodity SSDs},'' in \emph{Proc. of USENIX FAST}.\hskip 1em plus 0.5em minus 0.4em\relax USENIX Association, Feb. 2015, pp. 45--58.

\end{thebibliography}
%%%%%%%%%%%%%%%%%%%%%%%%%%%%%%%%%%%%

\end{document}